\documentclass[10pt]{iopart}

\usepackage{graphicx}
\usepackage{dcolumn}
\usepackage{bm}
\usepackage{braket}
\usepackage{color}
\usepackage{siunitx}
\usepackage{float}
\usepackage{hyperref}
\usepackage{cite}
\usepackage{lineno}
\usepackage{multirow}
\usepackage{tabularx}
\usepackage{enumitem}

\begin{document}

\title{Diamond quantum thermometry: From foundations to applications}

\author{Masazumi Fujiwara}
\address{
Department of Chemistry, Graduate School of Natural Science and Technology, Okayama University, 3-1-1 Tsushimanaka, Kita-ku, Okayama 700-8530, Japan}
\address{Department of Chemistry, Graduate School of Science, Osaka City University, Sumiyoshi-ku, Osaka, 558-8585, Japan}
\ead{masazumi@osaka-cu.ac.jp}

\author{Yutaka Shikano}
\address{Graduate School of Science and Technology, Gunma University, 4-2 Aramaki, Maebashi, Gunma 371-8510 Japan}
\address{Quantum Computing Center, Keio University, 3-14-1 Hiyoshi, Kohoku, Yokohama, 223-8522, Japan}
\address{Institute for Quantum Studies, Chapman University, 1 University Dr., Orange, CA 92866, USA}
\address{JST PRESTO, 4-1-8 Honcho, Kawaguchi, Saitama 332-0012, Japan}
\ead{yshikano@gunma-u.ac.jp}

\vspace{10pt}
\begin{indented}
\item[]April 2021
\end{indented}

\begin{abstract}
    Diamond quantum thermometry exploits the optical and electrical spin properties of colour defect centres in diamonds and, acts as a quantum sensing method exhibiting ultrahigh precision and robustness. 
    Compared to the existing luminescent nanothermometry techniques, a diamond quantum thermometer can be operated over a wide temperature range and a sensor spatial scale ranging from nanometres to micrometres. 
    Further, diamond quantum thermometry is employed in several application, including electronics and biology, to explore these fields with nanoscale temperature measurements. 
    This review covers the operational principles of diamond quantum thermometry for spin-based and all-optical methods, material development of diamonds with a focus on thermometry, and examples of applications in electrical and biological systems with demand-based technological requirements. 
\end{abstract}

%
%
%
%
\ioptwocol

\section{Introduction}
Temperature is recognized as a macroscopic quantity in thermodynamics. However, measuring temperature is defined locally because all thermometers have a local contact probe. Nanotechnology-based nanoscale thermometry has led to a number of experimental observations that cannot be understood based on the simple macroscale thermodynamics. This prompted theoretical developments on nano- and microscale thermodynamics~\cite{hill,sekimoto} and an intense debate on the interpretation of the heterogeneous temperature distribution in living cells~\cite{suzuki2020challenge,Ghonge_2018}. 

A diamond quantum sensor is a potential high-precision nanoscale thermometer based on the quantum-enhanced sensing technology originating from quantum physics. This was initiated after the first demonstration of diamond quantum magnetometry, which exploited the sensitivity of nitrogen-vacancy (NV) spin systems to environmental magnetic fields~\cite{Balasubramanian2008,maze2008nanoscale}. 
The demonstration of diamond quantum thermometry, using NV centres, was reported in 2013 by three different groups with suggested applications in temperature control in microfluidics~\cite{Toyli8417}, spatial mapping of joule heating around thin metallic wires~\cite{neumann2013high}, and temperature measurements of living cells~\cite{kucsko2013nanometre}. 
Quantum thermometry emerged as a result of the quantum physicists attempts to realize immobilized atomic systems, such as molecules, quantum dots, and colour defect centres, in solids. Such immobilized atomic systems can exhibit atomic-grade quantum properties, including lifetime-limited optical resonance~\cite{Hwang2009,Siyushev2014,PhysRevLett.109.033604,Schroeder2017,PhysRevX.1.011007,Najer2019,Huber:20}, extremely stable fluorescence emission~\cite{Chipaux2018small,hui2019carbon,PhysRevApplied.12.014042}, and degree of freedom in spin manipulations with long spin coherence time ~\cite{Herbschleb2019,Bar-Gill2013,Pingault2017,PhysRevB.96.081201}, which can be excellent test beds for experiments on quantum optics, quantum information science, and quantum metrology~\cite{Schroder:16}.

This review aims to provide an overview of the technological developments in diamond quantum thermometry, in terms of its fundamental operational principles to real applications in materials/device science and biological science.
The real applications of diamond quantum thermometry require different quantum technological developments. For example, in standard laboratory quantum physics experiments, most experiments are performed in a stable, well-controllable environment, wherein the ideal sensitivity of the quantum sensors can be obtained. In contrast, a diamond quantum sensor is expected to have a high temperature sensitivity, owing to the quantum-enhanced technology. This sensitivity is limited by the quantum properties of the NV centres, and approximately 10 mK/$\sqrt{\rm Hz}$ was demonstrated using single NV centres in pure bulk diamonds~\cite{Toyli8417,kucsko2013nanometre}.
In addition to temperature sensitivity, the diamond quantum thermometers exhibit fluorescence stability and a large operational temperature range of 100--1000 K~\cite{PhysRevApplied.12.014042,liu2019coherent}. The chemical inertness of diamond enables its use in harsh physicochemical environments or artefact-free sensing in complex biological environments.
While the abovementioned sensitivity in bulk diamonds is degraded to 0.1--5 K/$\sqrt{\rm Hz}$ in nanodiamonds (NDs), i.e., by a factor of 10--100, because of the deteriorated quantum properties of the NV centres, these promising properties have driven its technological development very rapidly, thereby extending its applications to more practical cases, such as the characterization of electronic devices and the precise temperature probing of cells and small animals.

As mentioned above, diamond quantum thermometry is one of the candidate element technology of nanoscale thermometry. Compared to other nanoscale thermometry techniques, such as thermo-responsive molecular probes~\cite{doi:10.1021/ac901644w,doi:10.1021/ac010370l,kriszt2017optical,SUZUKI2007L46,Bradac2020all-optical,okabe2012intracellular}, quantum dots~\cite{doi:10.1021/nn201142f,santos2018vivo,del2018vivo}, and Raman spectroscopy~\cite{doi:10.1002/anie.201915846,974795,1705092}, it has several advantages, including fluorescence stability, larger operational range, robust sensing, and high-temperature sensitivity. 
However, the diamond quantum thermometry has some drawbacks as well. For example, introduction of unnecessary nanoparticles into the specimen (e.g., electronic devices and living cells) may limit the versatility of the applications, as compared to near-infrared (NIR) thermography, Raman thermometry, and gene-coded fluorescent protein thermometry~\cite{kiyonaka2013genetically}. 
One of the crucial objectives of this review is to highlight and discuss these technical issues of diamond quantum thermometry, which need to be solved in the future. However, we do not intend to provide a comparative analysis between the various luminescence nanothermometry techniques, as provided in several other reports and reviews~\cite{Bradac2020all-optical,QUINTANILLA2018126,suzuki2020challenge,Zhou2020,glushkov2019fluorescent,bednarkiewicz2020standardizing}. 

\begin{figure*}[th!]
 \centering
 \includegraphics[scale=1.0]{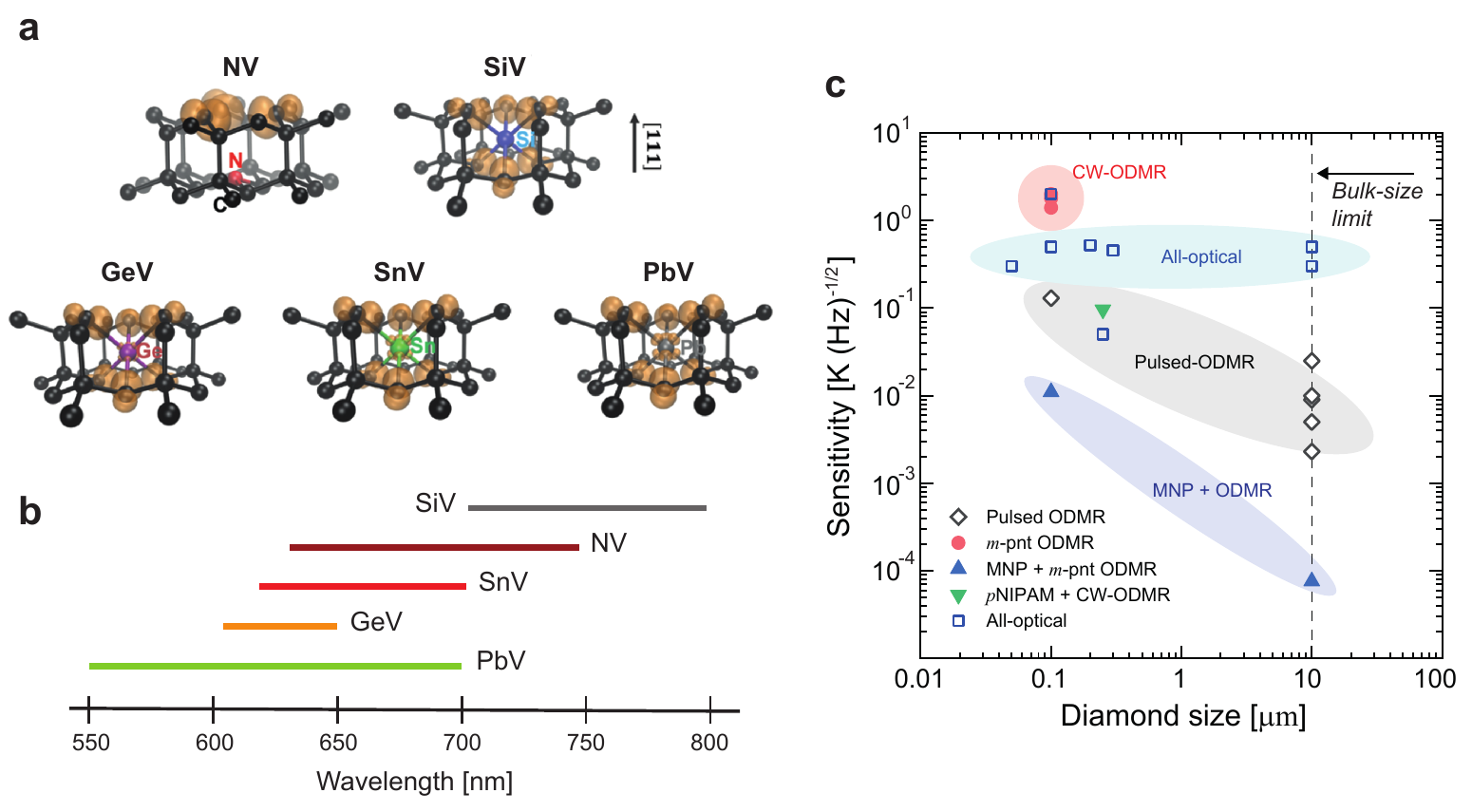}
 \caption{(a) Structures and spin polarization densities of spin defects in diamond for NV centres and group-IV colour centres (SiV, GeV, SnV, and PbV) (from \cite{he2020PCCP}) and (b) their fluorescence spectral range.
 (c) Plots of temperature sensitivity vs. diamond size; data summarized in Table~\ref{tbl1} with references. Notably, the sensitivity is dependent on the fluorescence photon count (or the optical excitation intensity). This brightness factor is not compensated in this figure.
 }
\label{fig1}
\end{figure*}

\renewcommand{\arraystretch}{1.1}
\begin{table*}[th!] 
  \caption{Size and sensitivity of diamond quantum thermometry, categorized by measurement methods. ZPL: zero-phonon line.}
  \label{tbl1}
  \centering
   \begin{tabular}{p{20mm}p{20mm}p{35mm}p{30mm}p{15mm}p{10mm}}
    \hline \hline
	\multirow{2}{*}{Sensor size} & Sensitivity  &\multirow{2}{*}{Method} & \multirow{2}{*}{Form} &\multirow{2}{*}{Defect}				& \multirow{2}{*}{Ref.}\\
	  & $({\rm mK} / \sqrt{\rm Hz})$	&  &   & \\
    \hline \hline
    \multicolumn{3}{l}{Spin-based thermometry} & \\
    	1 mm & 25	& Pulsed ODMR	&   Bulk		& NV		&  \cite{Toyli8417} \\
	1 mm & 5.0	& Pulsed ODMR	&   Bulk		& NV		&  \cite{neumann2013high} \\
	1 mm & 9.0	& Pulsed ODMR	&  Bulk		& NV		&  \cite{kucsko2013nanometre} \\
	1 mm & 10	&  Pulsed ODMR	&    Bulk		& NV		&  \cite{PhysRevB.91.155404} \\
	3 mm & 2.3	& Pulsed ODMR	&    Bulk		& NV			&  \cite{clevenson2015broadband} \\
	100 nm & $1.3 \times 10^2$	&  Pulsed ODMR	&  ND  & NV				&  \cite{neumann2013high} \\
	100 nm & $1.8 \times 10^3$	& m-pnt CW ODMR	&   ND  & NV					&  \cite{kucsko2013nanometre,choi2019} \\
	100 nm & $2.0 \times 10^3$	& m-pnt CW ODMR	&   ND  & NV					&  \cite{tzeng2015time} \\
	100 nm & $1.4 \times 10^3$	& m-pnt CW ODMR	&   ND  & NV					&  \cite{fujiwara2020realtime} \\
	100 nm & 1.1	& m-pnt CW ODMR	&   ND + MNP	& NV			&  \cite{PhysRevX.8.011042} \\
	1 mm & $7.6 \times 10^{-2}$ & m-pnt CW ODMR	&  Bulk-NP + MNP	&NV			&  \cite{Liu2020} \\
	250 nm & 96	& CW ODMR	&  ND@pNIPAM & NV				&  \cite{doi:10.1021/acsphotonics.9b00468} \\
	\hline
	\multicolumn{3}{l}{All-optical thermometry} & \\
	250 nm & 50 (or 13)	& multiparametric	&  ND  & SiV		&  \cite{kucsko2013nanometre} \\
	50 nm & $3.0 \times 10^2$	& ZPL intensity	&  ND  & NV					&  \cite{Plakhotnik_2015} \\
	1 mm & $3.0 \times 10^2$	& ZPL intensity, shift	&  Bulk  & GeV				&  \cite{doi:10.1021/acsphotonics.7b01465} \\
	200 nm & $5.2 \times 10^2$	& ZPL shift	&  ND & SiV					&  \cite{nguyen2018alloptical} \\
	300 nm & $4.6 \times 10^2$	& Antistokes	&  ND &  SiV				&  \cite{Traneaav9180} \\
	1 mm & $5.0 \times 10^2$	& ZPL linewidth, shift	&  Bulk & SnV			&  \cite{doi:10.1063/1.5037053} \\
	100 nm & $2.0 \times 10^3$	& ZPL shift	&  ND  & NV					&  \cite{tsai2017measuring} \\
	100 nm & $5.0 \times 10^2$	& ZPL intensity, shift	&  ND  & NV				&  \cite{Hui2019} \\
    \hline \hline
  \end{tabular}
\end{table*} 
\renewcommand{\arraystretch}{1.0}

This paper is organized as follows: 
In section~\ref{sec2}, we describe the types of diamond colour centres and thermometry methods (Sec.~\ref{sec2-0}) with the measurement principles of diamond quantum thermometry for spin-based (Sec.~\ref{sec2-1}) and all-optical thermometry (Sec.~\ref{sec2-2}). 
For spin-based thermometry, we focus on NV centres as only the NV centres can provide a practical level of detection at room temperature, whereas the optical detection of spins has been observed in other colour centres~\cite{Bradac2019quantum}. 
For all-optical thermometry, we report both NV centres and other group-IV colour centres as promising candidates.
In section~\ref{sec2-image}, optical detection method is discussed as it strongly affects the measurement strategy of diamond quantum thermometry. 
In section~\ref{sec2-mw}, methods of microwave delivery in the spin-based thermometry is described with a focus on microwave antenna structures.
In section~\ref{sec2-3}, methods to determine experimental sensitivity are discussed. 
The morphological form of diamond thermometers---NDs and bulk diamonds---is described in section~\ref{sec2-4}. 
In section~\ref{sec3}, we focus on the material properties of diamonds, which are particularly related to thermometry applications. 
The surface preparation of diamonds is described in Sec.~\ref{sec3-1}. 
In section~\ref{sec3-2}, we describe the temperature dependence of the key material properties used as a temperature indicator in thermometry. 
In section~\ref{sec3-3}, the robustness and possible artifacts of the temperature indicators in the thermometry are discussed.
In section~\ref{sec4}, we describe the application of diamond quantum thermometry with an emphasis on the future directions of research and development. 
Following the general introduction on the significance of nanoscale thermometry in electronic device characterization (section~\ref{sec4.1}), we show examples of temperature measurements of nanoscale electronic devices using diamond quantum thermometry in section~\ref{sec4.2}. 
In section~\ref{sec4.3}, we describe the use of diamond quantum thermometry for thermal conductance measurements.
In section~\ref{sec5.1}, we first discuss the significance of temperature probing at the nanoscale for the introduction of diamond quantum thermometry into biological applications. 
We introduce the experiments of diamond quantum thermometry in combination with photothermal laser heating in section~\ref{sec5.2}. 
The use of diamond quantum thermometry to monitor biological activities in various scenarios in section~\ref{sec5.3}. 
In section~\ref{sec6}, we conclude and discuss the remaining challenges for diamond quantum thermometry.

\section{Diamond quantum thermometry methods}
\label{sec2}

\subsection{Types of colour centres and thermometry methods}
\label{sec2-0}
Nitrogen and group-IV colour defect centres in diamond, including nitrogen-, silicon-, germanium-, tin-, and lead-vacancy centres (NV, SiV, GeV, SnV, and PbV), exhibit distinct optical and spin properties that can be used for various quantum applications, such as the development of single photon sources~\cite{Babinec2010,schroder2012nanodiamond,doi:10.1021/nl103434r,fujiwara2017fiber,Iwasaki2015,Rogers2014}, quantum gates~\cite{PhysRevB.95.205420,Nagata2018}, quantum memories~\cite{Tsurumoto2019,PhysRevLett.119.223602}, and quantum sensors~\cite{RevModPhys.89.035002,https://doi.org/10.1002/qute.202000066}. Reportedly, these colour centres can be used as thermometers, and have a wide fluorescent wavelength range and spin properties, as shown in Fig.~\ref{fig1}a, b. 
Based on their temperature-dependent properties, the thermometry methods are classified into spin-based thermometry and all-optical thermometry. 
The NV centres are used in both spin-based and all-optical thermometry at room temperature, whereas the other colour centres are primarily used in all-optical thermometry. The sensitivities and diamond sizes of the previous reports are summarized in Fig.~\ref{fig1}c and Table~\ref{tbl1}. 
In the rest of this review, we mainly focus on the NV diamond quantum sensors.

\subsection{Spin-based thermometry}

\label{sec2-1}
\begin{figure}[t!]
 \centering
 \includegraphics[scale=1.0]{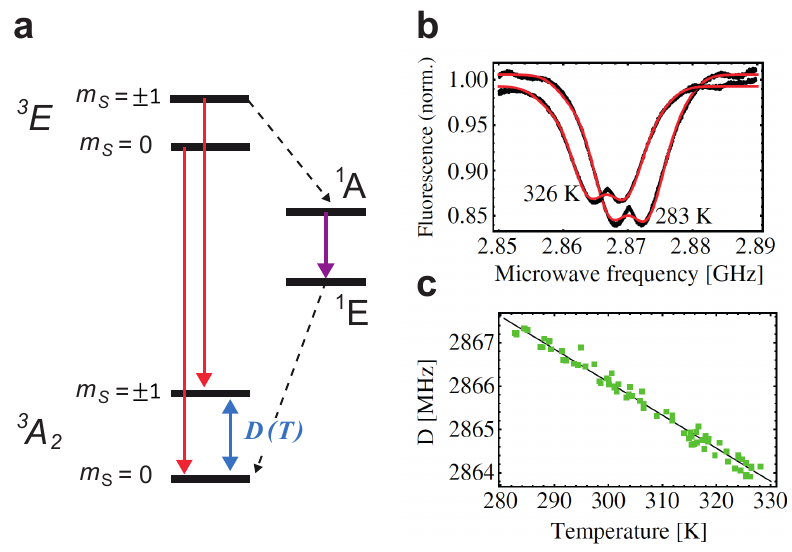}
 \caption{ (a) Energy diagram for NV centres. $D(T)$ indicates the temperature-dependent zero-field splitting. The red and violet arrows indicate the fluorescence between the triplet states ($^3E$ $ \rightarrow$ $^3A_2$) and a weak NIR fluorescence between the singlet states ($^1A$ $\rightarrow$ $^1E$), respectively. The dashed arrows represent nonradiative relaxation pathways. (b) ODMR spectra of the NV centres in bulk diamonds at 283 K and 326 K, and (c) the associated temperature dependence of $D(T)$ (from~\cite{PhysRevLett.104.070801}).
 }
 \label{fig-meth-cwodmr}
\end{figure}

\subsubsection*{\textsf{Principles of spin-based thermometry}}
Diamond quantum thermometry is distinct from other optical thermometry techniques because of its use of electron spin resonance to measure the temperature, particularly using the NV centres. 
The NV centres are spin-1 systems ($S = 1$) under the $C_{3v}$ symmetry and have triplet ground states ($^3A_2$) with electron spin sublevels of $m_S = 0$ and $m_S = \pm1$. Their zero-field splitting (ZFS) frequencies are at approximately $D = 2.87$ GHz (Fig.~\ref{fig-meth-cwodmr}a). 
The NV spin state is initialized to $m_S = 0$, under optical excitation, because the electrons in $m_S = \pm1$ of the excited state ($^3E$) undergo intersystem crossing to $m_S = 0$ of the ground state via the associated singlet states ($^1A$, $^1E$).
By applying the microwave resonance to the transition between $m_S = 0$ and $m_S = \pm1$ in the ground state, the fluorescence decreases substantially---a process called the optically detected magnetic resonance (ODMR) (Fig.~\ref{fig-meth-cwodmr}b). 
$D$ is affected by the magnetic field as well as the electric field, temperature, and pressure. These characteristics have recently attracted significant attention for quantum sensing using magnetometry, electrometry, and thermometry. 
Thermometry uses the temperature dependence of $D$, originating from the thermal lattice expansion and temperature dependence of the
electron--phonon interaction~\cite{PhysRevB.90.235205,PhysRevB.90.041201,doi:10.1063/1.3652910}. Although this theoretical mechanism has been explained by a phenomenological approach, it has not yet been understood as a microscopic model of the NV centre spin Hamiltonian~\cite{Gali2019}. This temperature dependence was first studied in the bulk diamonds in the range of 280--330 K by Acosta et al. ~\cite{PhysRevB.90.041201,PhysRevLett.104.070801} (Fig.~\ref{fig-meth-cwodmr} (c). 
The dependence in the lower temperature range of 5.6--295 K was later studied by Chen et al. ~\cite{doi:10.1063/1.3652910}. 
Following these reports, quantum-enhanced high-temperature sensitivity of diamond NV centres were demonstrated using sophisticated quantum spin control protocols~\cite{Toyli8417,kucsko2013nanometre,neumann2013high}. 
With this wide range of temperature responsivity, spin-based thermometry has been demonstrated to work in the range of 150--1000 K~\cite{doi:10.1063/1.3652910,PhysRevX.2.031001,liu2019coherent}.
Three types of measurement protocols can be used to read the temperature-dependent frequency shift of ODMR in the NV centres: (i) continuous-wave (CW) ODMR spectral measurement method, (ii) multipoint ODMR method, and (iii) pulsed ODMR method. 

\subsubsection*{\textsf{CW-ODMR method}}
\begin{figure}[th!]
 \centering
 \includegraphics[scale=1.0]{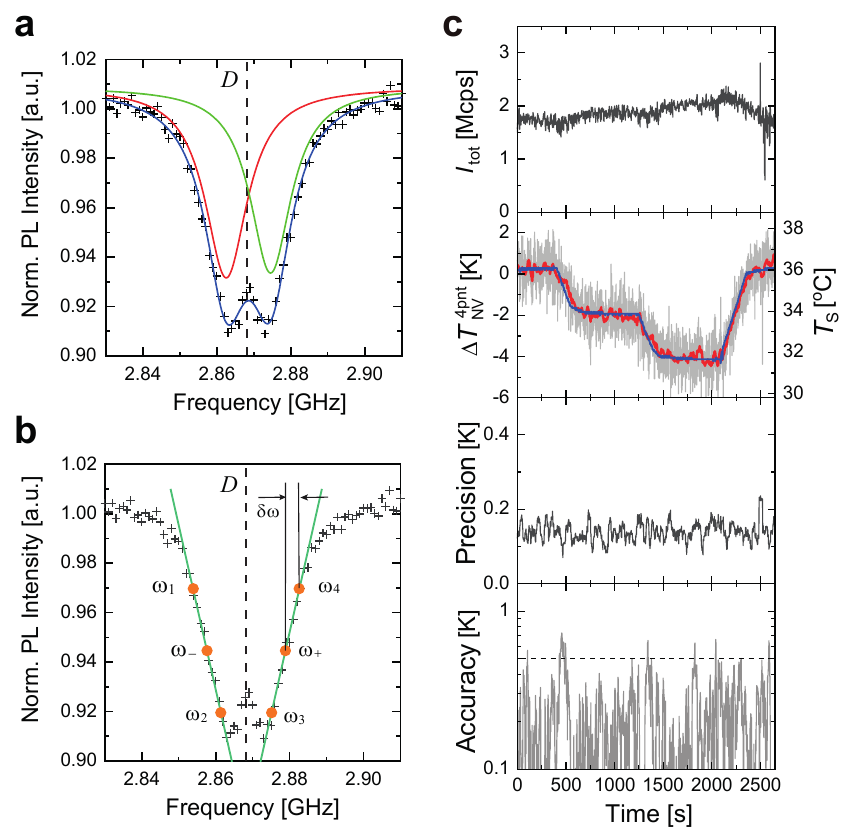}
 \caption{(a) ODMR spectrum fitted using the sum of two Lorentzian functions and (b) with two linear slopes fitted to the spectrum. The frequencies $\omega_1$ to $\omega_4$, indicated by orange circles, are used for the four-point method. $\omega_-$ and $\omega_+$ are used for the three-point method and are located on each of the slopes by $\delta \omega$ ($\omega_{j} = \omega_\mp \pm \delta \omega$). (c) Time profiles of total fluorescence intensity ($I_{\rm tot} = \sum I_j$) (top), $\Delta T_{\rm NV}^{\rm 4pnt}$ (second top), precision (second bottom), and accuracy (bottom) for the four-point measurements. In the second top panel, gray, red, and blue indicate 1 s sampling data, 20-point moving average, and sample temperature $T_{\rm S}$ calibrated separately, respectively (from \cite{PhysRevResearch.2.043415}). 
 }
 \label{fig-m-point}
\end{figure}

The CW-ODMR method is the most straightforward approach to measure the frequency shift of the ZFS; this method acquires the entire spectral shape of the ODMR at two different temperatures. The temperature sensitivity can be calculated using~\cite{RevModPhys.92.015004,liu2019coherent}: 
\begin{equation}
    \eta_{T} = \frac{\Delta \omega}{C \sqrt{R} |dD/dT|}, 
    \label{eq:1}
\end{equation}
where $C$, $\Delta \omega$, $R$, and $dD/dT$ denote the ODMR contrast, ODMR linewidth, detected photon rates, and temperature dependence of the ZFS, respectively.
The major advantage of using the CW-ODMR method is the wide operational range of the microwave frequency, which becomes important when the measurements are performed in a wide temperature range.
Additionally, the methodological simplicity of this method is also important, as it is usually applied, prior to the multipoint or pulsed ODMR methods, to find what the resonance looks like. 
Although this method is simple and yields significant information on the spectral changes between the spectra acquired at two different temperatures, the measurement is substantially slow for most real applications.
Furthermore, it is challenging to reach the theoretical limit of measurement precision for NDs containing ensemble NV centres because of the presence of a split structure in the ODMR dip. This dip in the spectra appears due to the crystal strain~\cite{tetienne2013spin,fujiwara2016manipulation} or spin-state interference of the NV centres~\cite{matsuzaki2016optically}. 
These split structures inevitably increase the error in locating the central frequency by curve fitting of a peak profile with any shape, including Lorentzian and Gaussian (see Fig.~\ref{fig-m-point}a), resulting in a fitting error up to 200--300 kHz (several Kelvin), even after performing data acquisition for several minutes.
Consequently, the practical sensitivity becomes worse than the theoretical sensitivity, presented in Eq.~\ref{eq:1}.
Some practical methods have been proposed to avoid this intrinsic error while estimating the spectral shape using analytical functions. These methods include the removal of the split region from the fitting data ~\cite{tzeng2015time,simpson2017non} or fitting the ODMR shape with fifth-order polynomial functions~\cite{tzeng2015time}.

\subsubsection*{\textsf{Multipoint ODMR method}}
In principle, the multipoint ODMR method is the CW-ODMR method, albeit with significantly enhanced measurement speed and practical precision. In this method, the ODMR shift is determined by measuring only the fluorescence intensity at several frequency points. Schoenfeld demonstrated this efficient speed-up protocol, with two frequency points, as real-time magnetometry~\cite{PhysRevLett.106.030802}. This approach was further extended to the symmetric four-point method using an effective noise cancelling approach by Kucsko et al. ~\cite{kucsko2013nanometre}. In this four-point method, the temperature can be determined by 
\begin{equation}
\Delta T_{\rm NV}^{\rm 4pnt} = \frac{\delta \omega}{dD/dT} \frac{(I_1+I_2) -(I_3+I_4)}{(I_1-I_2) - (I_3-I_4)}, 
\end{equation}
where $I_j$ indicates the fluorescence intensity at the four frequency points ($\omega_j$). 
$\delta \omega$ is defined as $\omega_{j} = \omega_\mp \pm \delta \omega$ (Fig.~\ref{fig-m-point}a, b).
Later, Tzeng et al. proposed a three-point method by assuming that the ODMR spectral shape is a single Lorentzian; this method has the advantage of reducing the required microwave components~\cite{tzeng2015time,PhysRevX.8.011042}. 

These multipoint ODMR methods accelerate the overall measurement speed for temperature estimation and allow a large amount of data integration within a short period of time, thus providing high precision in real time.
The experimentally determined temperature sensitivity becomes approximately 2.0 ${\rm K}/\sqrt{\rm Hz}$~\cite{tzeng2015time,fujiwara2020realtime} (Fig.~\ref{fig-m-point}c) and about 10 ${\rm mK} / \sqrt{\rm Hz}$, when exploiting the magnetic criticality effect of magnetic nanoparticles~\cite{PhysRevX.8.011042}.
The measurement speed and the possibility of incorporating artefacts is a trade-off  because of the intrinsic nature of the estimation process, which is based on a few fluorescence intensity values in the ODMR spectra. 
For example, the optical power dependence of the ODMR spectral shape causes an anticrossing shift in the resonance frequency of $\ket{0} \rightarrow \ket{\pm 1}$ for NDs containing an ensemble of NV centres~\cite{PhysRevResearch.2.043415}. 
This frequency shift becomes an artefact in the temperature measurements using the multipoint ODMR methods.

\begin{figure}[th!]
 \centering
 \includegraphics[scale=1.0]{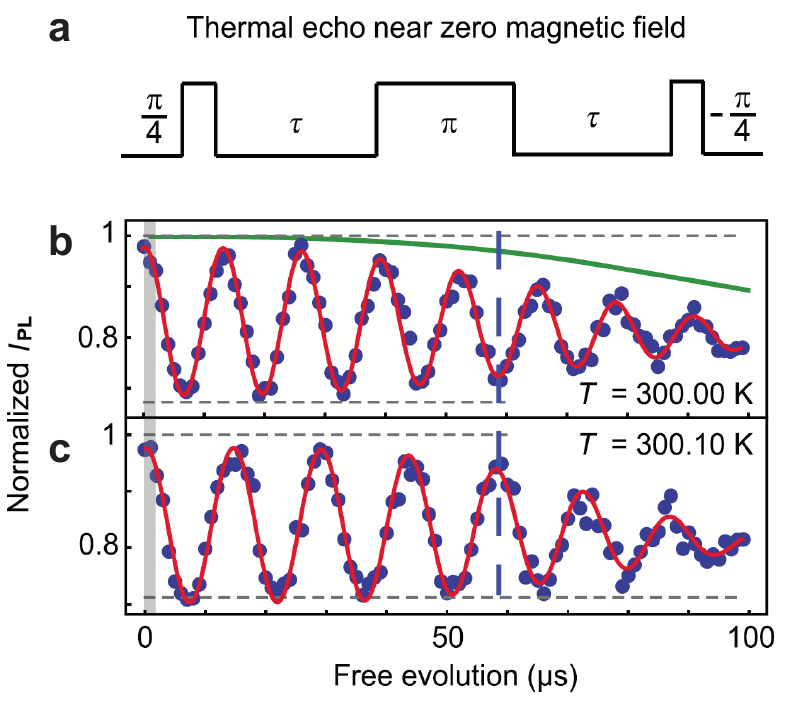}
 \caption{(a) Pulse sequence of thermal echo and (b) the resultant oscillations at two temperatures with NV centres in bulk diamonds (from \cite{Toyli8417}). 
 }
 \label{fig-pulseodmr}
\end{figure}

\subsubsection*{\textsf{Pulsed ODMR method}}
The pulsed ODMR methods provide high sensitivity by eliminating the decoherence effect of the spin systems. In this case, the sensitivity is expressed as: 
\begin{equation}
    \eta_{T} = \frac{1}{C |dD/dT| \sqrt{R t_{\rm coh}}}, 
    \label{eq:2}
\end{equation}
where $t_{\rm coh}$ denotes spin coherence time.
The previous pulsed ODMR methods were based on a thermal echo sequence; the corresponding microwave pulse sequence is shown in Fig.~\ref{fig-pulseodmr}a. 
In this sequence, the initial $\pi/4$ pulse creates a superposition of $\ket{0}$ and $\ket{\pm 1}$. After an evolution time of $\tau$, the $2 \pi$ pulse interchanges the population of the $\ket{+1}$ and $\ket{-1}$ states.
The temperature-induced frequency shift can be detected by following another period of free evolution for time $\tau$ with the second $- \pi/4$ pulse.
Using this protocol, a temperature sensitivity of 10 ${\rm mK} / \sqrt{\rm Hz}$ was demonstrated in bulk diamonds, whereas a sensitivity of 130 ${\rm mK} / \sqrt{\rm Hz}$ was reported for NDs using the D-Ramsey sequence (a variation of the thermal echo method)~\cite{neumann2013high}.
Konzelmann et al. reinforced the D-Ramsey model with optimal control theory and additional pulse sequences for a dynamic noisy environment. With this reinforced version, they demonstrated a stable temperature measurement with a sensitivity of sub-100 ${\rm mK} / \sqrt{\rm Hz}$ for rotating NDs in the agarose matrix. Notably, the magnetic-field alignment was used in the D-Ramsey model to improve the stability and sensitivity of the magnetometry~\cite{Abobeih2019}.
The advantage of using the pulsed ODMR method is that it's sensitivity is within 10--100 ${\rm mK} / \sqrt{\rm Hz}$, which is one order of magnitude smaller than those of the CW-ODMR method.
Conversely, the relatively long measurement times in the practical experiments may be a challenge.
In the pulsed ODMR experiments, overhead time associated with the spin measurements may significantly deteriorate the sensitivity and precision.
The ideal sensitivity described in Eq.~\ref{eq:2} cannot be achieved always in real experimentation, as recently described for the magnetometry case analysed by Barry et al.~\cite{RevModPhys.92.015004}.

\subsection{All-optical thermometry}
\label{sec2-2}
\begin{figure}[th!]
 \centering
 \includegraphics[scale=1.0]{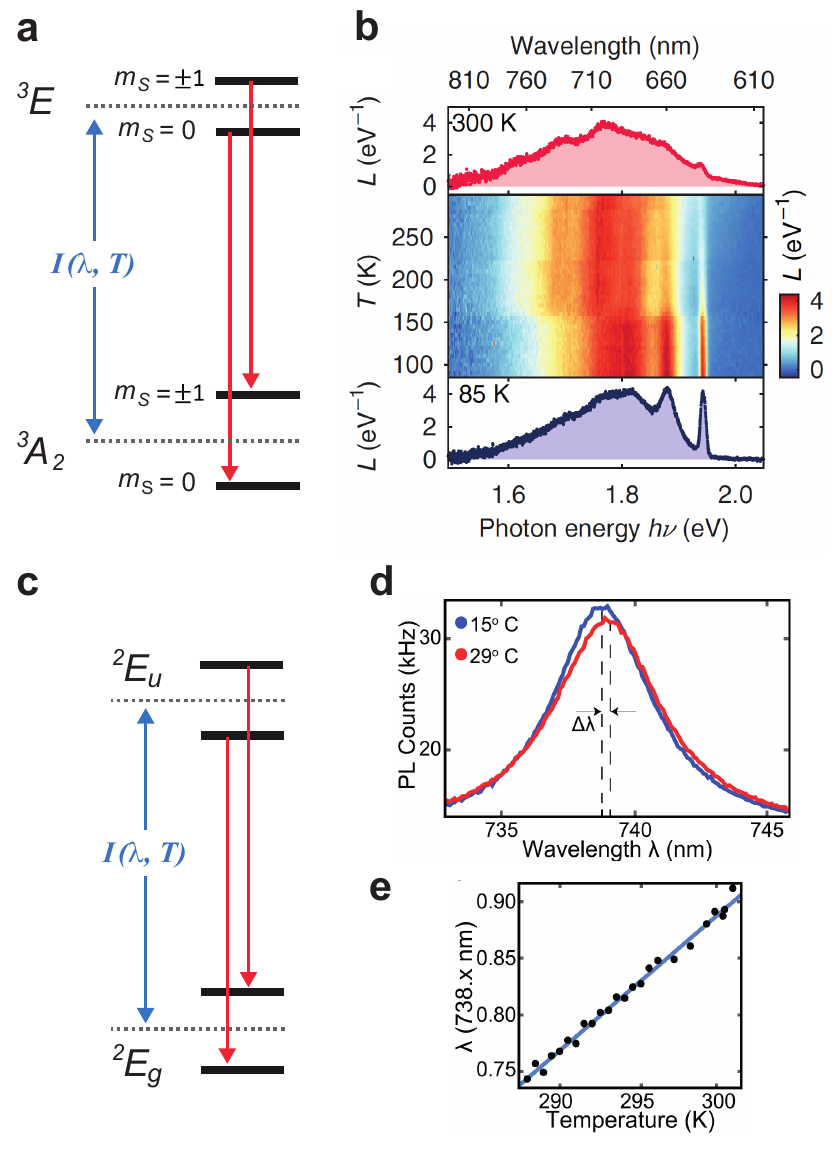}
 \caption{(a) Energy diagram of NV centres in all-optical thermometry. The red arrows indicate fluorescence, and $I (\lambda,\ T)$ indicates the temperature-dependent optical transition. (b) Evolution of the NV fluorescence spectrum between 85 K and 300 K.  Discontinuities at $T \sim 150$ and 230 K are associated with the phase transitions of the polymer in the devices (adapted from~\cite{PhysRevApplied.12.014042}). (c) Energy diagram of the group-IV colour centres in all-optical thermometry. (d) Typical fluorescence spectra of SiV centres at 15 $\si{\degreeCelsius}$ (blue) and 29 $\si{\degreeCelsius}$ (red). The ZPL shows redshift at higher temperatures. (e) ZPL position as a function of temperature adapted from~\cite{nguyen2018alloptical}).
 }
 \label{fig-allopt-meth}
\end{figure}

While spin-based thermometry promises high sensitivity and robustness, the necessity of microwave excitation can restrict its applications in areas or fields where the technical complexity of microwave power delivery or heating effect is a challenge. 
In these cases, the all-optical thermometry can be used as an alternative method, which exploits the temperature dependence of the fluorescence spectral shape (Fig.~\ref{fig-allopt-meth}a).
Plakhotnik et al. first proposed the use of temperature-dependent optical properties of the NV centres, such as the fluorescence intensity, lifetime, and spectral shape in the range of 300--700 K, for thermometry~\cite{plakhotnik2010luminescence}. 
The variability in the fluorescence intensity and lifetime were approximately $-$0.2 \%/K and  $-$0.06 ns/K, respectively. 
They proposed a thermometry based on the Debye--Waller factor of the NV fluorescence spectrum (i.e., the ratio of zthe ero-phonon line (ZPL) to the entire spectrum)~\cite{Plakhotnik2014all}.

Because the fluorescence intensity fluctuates in most experiments, employing the Debye--Waller factor ensures robustness of the temperature measurement against noise. 
This ratiometric approach provides a temperature sensitivity of 0.3 ${\rm K}/\sqrt{\rm Hz}$~\cite{Plakhotnik_2015}. 
The same measurement principle was applied to cryogenic thermometry, in the range from 80--300 K, by Fukami et al.~\cite{PhysRevApplied.12.014042} (Fig.~\ref{fig-allopt-meth}b).
They succeeded in observing a temperature gradient over tens of micrometres, in a ferromagnetic insulator substrate, at cryogenic temperatures. 
The temperature dependence of the fluorescence lifetime of the NV centres may be used for robust thermometry; however, it was not implemented until now.
In fact, fluorescence lifetime measurements have been used for other luminescent nanothermometers, such as thermo-responsive polymer nanoparticles~\cite{okabe2012intracellular}. 

The technical simplicity of the all-optical thermometry allows temperature measurements for various colour defect centres in diamonds, including the group-IV colour centres (SiV, GeV, SnV, and potentially PbV). 
These colour centres have $D_{3d}$ point symmetry, with ground ($^2E_g$) and excited ($^2E_u$) states of $E$ symmetry (Fig.~\ref{fig-allopt-meth}c)~\cite{Bradac2019quantum}. 
Each state has a pair of orbital sublevels, lifted by the spin--orbit coupling and the dynamic Jahn--Teller interaction. 
These spin structures cannot be used for thermometry at room temperature; therefore, the application of group-IV colour centres for thermometry was demonstrated only by the all-optical method. 
Nguyen et al. used the peak shift of the ZPL of fluorescence arising from SiV centres in NDs~\cite{nguyen2018alloptical}, which are less affected by the fluorescence intensity fluctuations (Fig.~\ref{fig-allopt-meth}d,e)~\cite{nguyen2018alloptical}.
The relatively slow measurement speed of the entire spectral shape can be accelerated by employing a variety of data-analysis methods. Choi et al. reported all-optical thermometry with SiV centres and multiparametric data analysis of the entire fluorescence spectral shape~\cite{doi:10.1021/acsphotonics.9b00468}. 
The fluorescence spectra, originating from the SiV centres, exhibit a temperature dependence that can be modelled using a characteristic spectral function with two Lorentzian peaks. 
The multiple parameters of this spectral function are determined by fitting the function to the experimental spectra with the aid of a multiparametric data analysis software. 
This approach demonstrated a temperature sensitivity of 13 ${\rm mK} / \sqrt{\rm Hz}$ (50 ${\rm mK} / \sqrt{\rm Hz}$ for a biologically realistic excitation intensity of 70 ${\rm kW}/{\rm cm}^{2}$)~\cite{doi:10.1021/acsphotonics.9b00468}. 
The all-optical method has been also applied to GeV~\cite{doi:10.1021/acsphotonics.7b01465,doi:10.1021/acsphotonics.9b00206} and SnV~\cite{doi:10.1063/1.5037053} centres. 
Recently, a new type of diamond quantum thermometry that employs the anti-Stokes excitation of the diamond defect centres has been reported~\cite{Traneaav9180}. 

The sensitivity values of the all-optical thermometry exhibit a size-independent trend, in contrast to those of the spin-based thermometry (see Fig.~\ref{fig1}c). 
In the spin-based thermometry, the sensitivity is degraded as the diamond size decreases owing to the surface effect of the NV spin systems, whereas this size-dependency is not observed in the all-optical method.
This indicates that the sensitivity of the all-optical thermometry is not limited by the colour centre's properties that are sensitive to the surface effect in the NDs.
While the physical reason of this trend is not clear, it might be related to the parameters of the optical instruments, such as wavelength resolution and sampling points. 
The size independence of the all-optical thermometry using diamonds may be understood in comparison with the other fluorescent nanothermometers, such as dyes, thermo-responsive polymers~\cite{okabe2012intracellular,uchiyama2015cationic,tsuji2017difference}, quantum dots~\cite{doi:10.1021/nn201142f,doi:10.1021/nn201142f,santos2018vivo,del2018vivo}, and rare-earth-ion nanoparticles~\cite{doi:10.1021/acs.inorgchem.9b00646,wang2015optical,WANG2021156744}, as they have common measurement principles. 
Other than the sensitivity, the real applications of each of these nanothermometers have different pros and cons, including fluorescence brightness, wavelength selectivity, and site specificity, as reported in recent review articles~\cite{okabe2018intracellular,Bradac2020all-optical}.

\subsection{Optical detection and imaging capability}
\label{sec2-image}
Both of the spin-based and all-optical diamond quantum thermometry rely on the detection of the fluorescence, and therefore optical detection methods critically affect the sensing performance and strategy.
There are two optical detection schemes; one is a photodetector-based single-point detection~\cite{Toyli8417,neumann2013high,kucsko2013nanometre,tzeng2015time,PhysRevX.8.011042,fujiwara2020realtime} and the other is a camera-based wide-field detection~\cite{simpson2017non,sekiguchi2018fluorescent,sotoma2018enrichment,Sotomaeabd7888,Nishimura2021} schemes
The photon count rate significantly influences the temperature sensitivity ($\sqrt{R}$ in Eqs.~\ref{eq:1} and~\ref{eq:2}).
As long as the number of photons are detected at the same level of photon counts, both detection schemes provide comparable sensitivity, as analysed in the spin-based CW-ODMR method using the NV centres~\cite{Nishimura2021}.
Technically, single-point detection is often used in confocal microscopy to improve the spatial resolution and signal-to-noise ratio. Further, the use of pinhole limits the detectable number of photons.
In contrast, the camera-based wide-field detection is able to detect greater number of photons; however, there are several associated factors, such as more background fluorescence and smaller dynamic range of camera than the confocal detection, which degrade the sensitivity and result in a sensitivity comparable to that of the confocal detection.
Notably, there are some variations of the optical setup, such as photodiode-based fluorescence detection for fast and highly sensitive spin measurements~\cite{levine2019principles} and super-resolution microscopy for gaining higher spatial resolution~\cite{Maurer2010,Jaskula:17}

The selection of optical detectors depends on the temperature sensitivity and spatial range of thermal probing necessary for the target applications. 
In the spin-based thermometry, the photodetector-based single-point detection is used in every method~\cite{kucsko2013nanometre,sotoma2017review}, and the camera-based wide-field detection has been demonstrated only in the CW-ODMR method~\cite{foy2019wide,simpson2017non,sotoma2018enrichment} because of the limited temporal resolution of the camera detectors.
In both detection schemes, increasing the temperature probing area with high temperature precision in a short measurement time is desirable. 
A current challenge for the single-point detection is the limited number of probable NDs in the microscopy field of view, while an increase in the temperature precision is required by the wide-field detection technique.
In the all-optical thermometry, only the single-point detection scheme has been demonstrated because of the necessity of fluorescence spectral measurements using spectrometers~\cite{Bradac2020all-optical} (the detectors themselves were camera, but they probed only a single point in a microscope image). 
Its extension is expected to be similar to Raman imaging~\cite{stewart2012ramanimaging}, in which a spectrometer equipped with fast high-sensitive camera are operated synchronously with spatial scanning.

\subsection{Microwave delivery in the spin-based thermometry}
\label{sec2-mw}
In the spin-based thermometry, microwave excitation at approximately 2.8 GHz is necessary to drive the NV spins.
This microwave frequency range is known as the S band. 
The microwave radiation is generated from a microwave source and its output timing is controlled using electrical switches or an arbitrary waveform generator via mixing~\cite{doi:10.1063/1.5011231,Dolde2013,Rosskopf2017}.
The timing controlled microwave amplitude is amplified and sent to an antenna. 
Different types of antenna designs are avaiable, including thin wire~\cite{neumann2013high}, coil~\cite{doi:10.1021/nl302979d,doi:10.1063/1.4952418}, coplanar waveguide~\cite{doi:10.1063/1.5028335,Sadzak2018}, omega shape antenna~\cite{Horowitz13493,choi2019,Miller2020} and loop-gap resonators~\cite{doi:10.1063/1.5037465,doi:10.1063/1.4952418,doi:10.1021/nl404072s}.
Historically, thin metallic wire was frequently used because of the technical simplicity and broad bandwidth having no resonance behaviour over several GHz; such a broad bandwidth is useful when measuring the ODMR spectra in a wide frequency range. 
However, such a thin wire structure cannot achieve impedance matching with coaxial cables (50 $\Omega$), and the insertion loss is more than 30 dB. 
In addition, spatial distribution of microwave magnetic field is highly heterogeneous (the field is concentrated around the wire).
The coil structure is another type of often-used antenna, as it provides spatially uniform field excitation, although the excitation power is not high. 
More recently, several loop-gap antenna structures have been proposed for efficient and spatially uniform microwave excitation in magnetometry. 
Bayat et al. reported a double split-ring resonator having approximately 25 MHz linewidth (quality factor of 120) and demonstrated efficient excitation of Rabi oscillation over an area of 1 $mm_2$ on bulk diamonds~\cite{doi:10.1021/nl404072s}. 
The narrow linewidth of the loop-gap resonators can be a problem while measuring the ODMR in a wide frequency range. 
Sasaki et al. reported a different type of loop-gap resonator with a bandwidth of 400 MHz. They designed the antenna to achieve efficient excitation of multiple ODMR lines as a compensation for the trade-off relation of the bandwidth and excitation efficiency~\cite{doi:10.1063/1.4952418}.
The loop-gap antenna structures have an additional advantage---the electric field is concentrated in the gap area and not in the sample area, which is important because it is crucial to avoid the effect of electric field on the sample for the real applications of diamond quantum thermometry.
The abovementioned studies were focussed on the magnetometry applications. 
The implementation of antenna structures in thermometry applications has not yet been reported.

\begin{figure}[t!]
 \centering
 \includegraphics[scale=1.0]{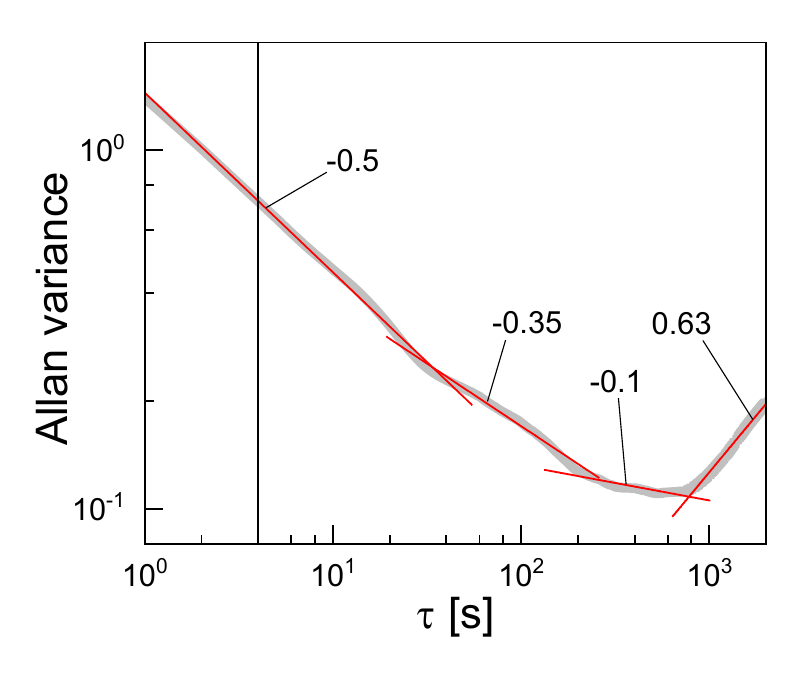}
 \caption{Allan variance profiles obtained from the steady temperature sensor signals as shown in the top panel in Fig.~\ref{fig-m-point}c. Slopes of the four distinct regions are indicated with values. The vertical line at 4 s indicates the tracking time used in this measurement (see Ref.~\cite{PhysRevResearch.2.043415} for details).
 }
 \label{fig-alavar}
\end{figure}

\subsection{Sensitivity determination}
\label{sec2-3}
In diamond quantum thermometry, measurement sensitivity is expressed as the temperature per square root of the frequency bandwidth.
Equations~\ref{eq:1} and \ref{eq:2} provide the theoretical sensitivity of NV-spin-based thermometry; a number of previous reports used this value as sensitivity. 
However, theoretical sensitivity may be different from experimental sensitivity exhibited by sensors in practice. As mentioned above, the split structure in ODMR of NV centers degrades the curve-fitting precision, and the experimental sensitivity is worse by a factor of 10, which indicates the importance of determining the sensitivity in real experiments. The experimental sensitivity ($\eta_{\rm exp}$) is the noise equivalent power of sensors, and it can be calculated by ~\cite{richards1994Bolometers} 
\begin{equation}
    \eta_{\rm exp} = \sigma_{\rm exp} \times \sqrt{2 \delta t_{\rm intgr}}, 
    \label{eq:3}
\end{equation}
where $\sigma$ and $\delta t_{\rm intgr}$ represent the precision values of the temperature dataset and integration time, respectively. These values can be obtained by recording the time profile of the temperature for a given time and by calculating the mean of the temperature data.

Measuring Allan variance data is important to determine the appropriate integration time~\cite{PhysRevX.5.041001,PhysRevB.97.024105,PhysRevResearch.2.043415,hatano2021APL}.
Figure~\ref{fig-alavar} shows the Allan variance data for the NV temperature sensor, used in Fig.~\ref{fig-m-point}c.
In this figure, there are four distinct regions of different slopes.
The first region with a slope of $-0.5$ is where the noise is governed by Gaussian noise.
The regions with slopes of $-0.1$ and $0.63$ are considered to originate from the temperature instability of the environment and the instability of the NV spin properties. 
The region with a slope of $-0.35$ is speculated to have originated from the mechanical instability of the microscope.
An important indication is that Eq.~\ref{eq:3} holds only in the Gaussian-noise region, with a slope of $-0.5$.
It is therefore reasonable to set the integration time ($t_{\rm intgr}$) to 40 s to achieve both high temperature precision and time resolution in the particular case shown in Fig.~\ref{fig-alavar}\cite{PhysRevResearch.2.043415}.

\subsection{Form of diamond sensor: Nanodiamonds and bulk diamonds}
\label{sec2-4}

\begin{figure}[t!]
 \centering
 \includegraphics[width=83mm]{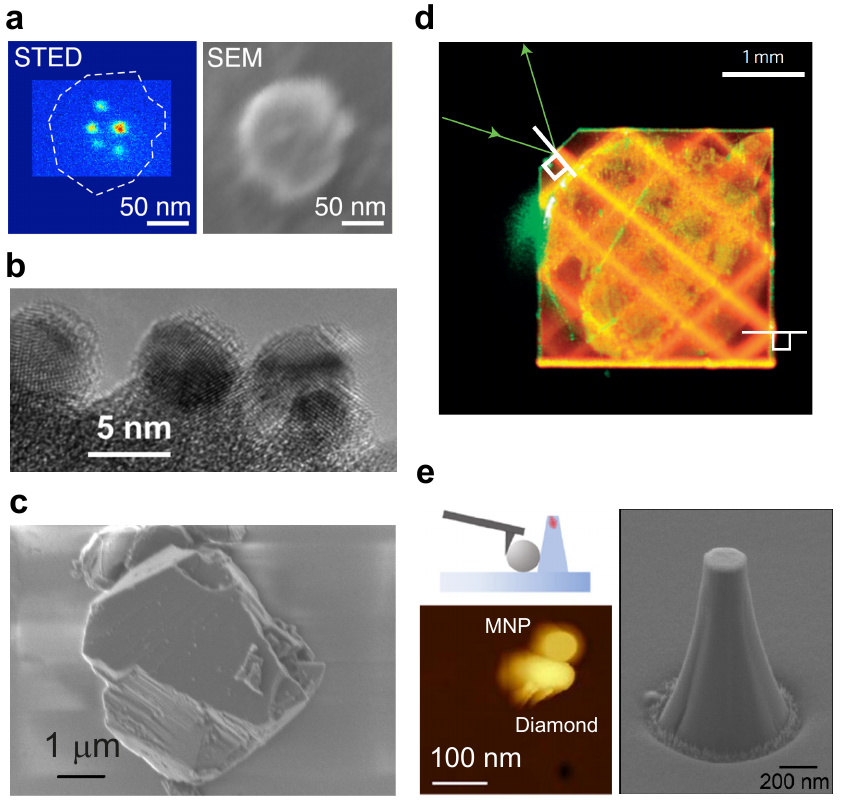}
 \caption{(a) Sub-diffraction STED image (left) and corresponding SEM image (right) of an ND particle, which possess five NV centers~\cite{doi:10.1021/nn404421b}. (b) High-resolution transmission electron microscopy (HRTEM) image of the primary particles of detonation ND purified from ND carbon by treatment in ozone~\cite{doi:10.1116/1.4927679}.
(c) SEM image of diamond microcrystals~\cite{doi:10.1063/1.5045535}. (d) Color image of a light-trapping diamond waveguide chip ($3\times 3 \times 0.3$ mm$^3$) excited by a green pump laser showing bright fluorescence without any spectral filtering~\cite{clevenson2015broadband}. (e) Schematic of the atomic force microscopy (AFM) nanopositioning of copper--nickel alloy magnetic nanoparticles (MNPs) to diamond nanopillars (left upper) and its AFM image (left bottom)~\cite{10.1093/nsr/nwaa194}. (right) SEM image of the diamond nanopillar.}
 \label{fig-diamondform}
\end{figure}

The form of the diamond that incorporates defect centers significantly affects its sensing ability. For NV centers, the best optical and spin properties are obtained for single NV centers in pure bulk diamonds, and the highest sensitivity was achieved with these systems~\cite{kucsko2013nanometre,Toyli8417,PhysRevB.91.155404}. 
However, contact thermometers need to consider how they themselves affect the temperature of the target objects via heat exchange at the physical interface between the target and thermometers. 
This is particularly the case for diamond nanothermometry, which aims to measure the temperature of tiny objects, because diamond has a very high thermal conductivity of 1000--3000 ${\rm W} / ({\rm m}\cdot \rm{K})$~\cite{doi:10.1063/1.5140030} (several times higher than that of copper).
Reducing the size of the diamond is necessary to probe the small heat source, whereas the significant size reduction degrades the defect centre properties. 
Further, the diamond chip size of tens of micrometres may not have a great advantage in terms of sensitivity compared to classical electronic thermometers such as thermopiles, wherein milli-Kelvin precision is readily available. 
Bulk diamonds were therefore only used for the benchmark studies, and most previous studies used NDs for nanothermometry applications.
The NDs can be distributed to various places including electronic devices and biological samples; they can sense the temperature at the locations once delivered. 
For electronic devices, NDs are either dropped onto or delivered with polymer films on circuit devices.
For most of biological applications, surface functionalization is necessary for site-specific targeting; as a result, the ND surface functionalization has been studied intensively ~\cite{mohan2010vivo,Mochalin2012,simpson2017non,hsiao2016fluorescent,haziza2017fluorescent,doi:10.1021/jp066387v,CHAUHAN20201}. 

The use of bulk diamonds for thermometry is relatively limited because of the trade-off between the achievable sensitivity and thermometer size. 
For example, at a sensor size of $\sim$0.1 mm, thermopile thermometers can achieve 0.27 mK$/\sqrt{\rm Hz}$ and be implemented successfully as microfluidic calorimeters~\cite{Lee15225}, which is one or two orders of magnitude higher than the best achievable diamond quantum thermometry.
However, there exist ongoing efforts to exploit the attractive high sensitivity of bulk diamonds for various practical applications.
Clevenson et al. reported a diamond chip device wherein the excitation laser is trapped via total reflection at the diamond/air interfaces to enhance the sensitivity of ODMR measurements~\cite{clevenson2015broadband}. 
By combining its sensing capability with multiple physical parameters, this chip device may be used as a compact precision sensor platform for measuring magnetic fields, temperature, pressure, rotation, or time.
Notably, the multiple quantum sensing ability of the diamond NV centres has been recently applied to characterize the phase change in lipid bilayers by exploiting spin-based magnetometry and thermometry. This represents another type of bulk-chip application~\cite{ishiwata2021label-free}. 

Nanostructured bulk diamonds have been investigated to introduce bulk-grade diamond quantum thermometry performance to local temperature probing. 
Tanos et al. reported nanopillar structures on bulk diamonds wherein NV centers show excellent spin properties~\cite{doi:10.1063/1.5140030}. 
They analyzed the heat conduction using the finite element method and evaluated the temperature sensing ability of these nanopillar structures. Wang et al. coupled such nanopillar structures with magnetic nanoparticles to boost the temperature sensitivity up to 11 mK$/\sqrt{\rm Hz}$ in real time, and they called it the critically enhanced magnetic sensitivity~\cite{PhysRevX.8.011042}. 
With an effective thermal management of the diamond nanostructures, one may harness the bulk-grade NV spin quality for thermometry.

\section{Material properties as thermometers}
\label{sec3}
\subsection{Surface preparation of diamonds}
\label{sec3-1}
The surface state strongly affects the properties of both bulk diamonds and NDs. 
Chemical termination of the diamond surface modifies the electronic band bending at the surface, resulting in a positive electron affinity, for oxygen and nitrogen termination, or negative affinity, for hydrogen termination~\cite{doi:10.1063/1.1900925,doi:10.1063/1.3561760,ROMANYUK2018208,doi:10.1063/1.4930945}. 
This band bending substantially influences the electrical conductivity of bulk diamonds. 
The spin and optical properties of the colour centres, and hence the thermometer performance, are also affected. 
For example, the charge state of the NV centres (NV$^0$, NV$^-$, NV$^+$) are determined in the balance between the band edge potential and the environmental potential, and the NV stability and spin coherence properties can be modified by the surface termination~\cite{PhysRevB.83.081304,Grotz2012,Schreyvogel2015,karaveli2016modulation,kawai2019Nitrogen,Lofgren2019thebulk}.
For diamond quantum thermometry applications, the oxygen or nitrogen terminated surface may be preferred because their electron positive affinity protects the colour defect centres' spin/optical properties.
The surface state also influences the chemical properties of the NDs~\cite{Arnault2015,TINWALA2019913,VILLALBA20111115,Zamani2020}, like the colloidal properties and surface reactivity towards further modification using additional functioning molecules.
The original surface of bare NDs is covered by various $sp^2$ carbon structures and possess multiple functional groups, such as C-H, -COOH, -COO and -OH.
Therefore, additional surface treatment, like oxygen plasma etching and strong acidic treatment, is required to attach additional organic molecules~\cite{wolcott2014surface,Nagl2015,tsukahara2019removing}.

To use NDs as thermometers, the NDs need to be delivered to the target places to measure the temperature.
For biological applications, some bioactive molecules are attached to the ND surface using these functional groups as a scaffold, including proteins and antibodies~\cite{doi:10.1021/la703482v,mohan2010vivo,Chang2019,doi:10.1021/acsami.6b15954,Miller2020,Cordina2018}. 
Some studies used polymeric structures (PG: polyglycerol, PEG: polyethylene glycol) to yield a higher dispersibility in buffer solutions with high ionic strengths, such as phosphate-buffered saline (PBS)~\cite{zhao2011chromatographic,ZOU2020395,zou2020Polyglycerol,MADAMSETTY2019112,madamsetty2019development,https://doi.org/10.1002/marc.201600344}.
These surface-functionalized NDs are delivered to target places by adding NDs to cell culture or via the mechanical delivery to samples through injection~\cite{Alkahtani:17,mohan2010vivo,choi2019,fujiwara2020realtime} or electropolation~\cite{Hemelaar2017,terada2018onepot}.
For electronic device applications, ND targeted delivery by surface functionalization has not yet been reported. 
According to the previous reports, the ND suspensions can be simply dropped into the target area in the devices.
%
%
Several review papers discuss the surface functionalization of NDs~\cite{doi:10.2217/17435889.4.1.47,doi:10.1002/anie.201905997,doi:10.1517/17425247.2015.992412}.

\subsection{Temperature dependence}
\label{sec3-2}
\begin{figure}[t!]
 \centering
 \includegraphics[width=70mm]{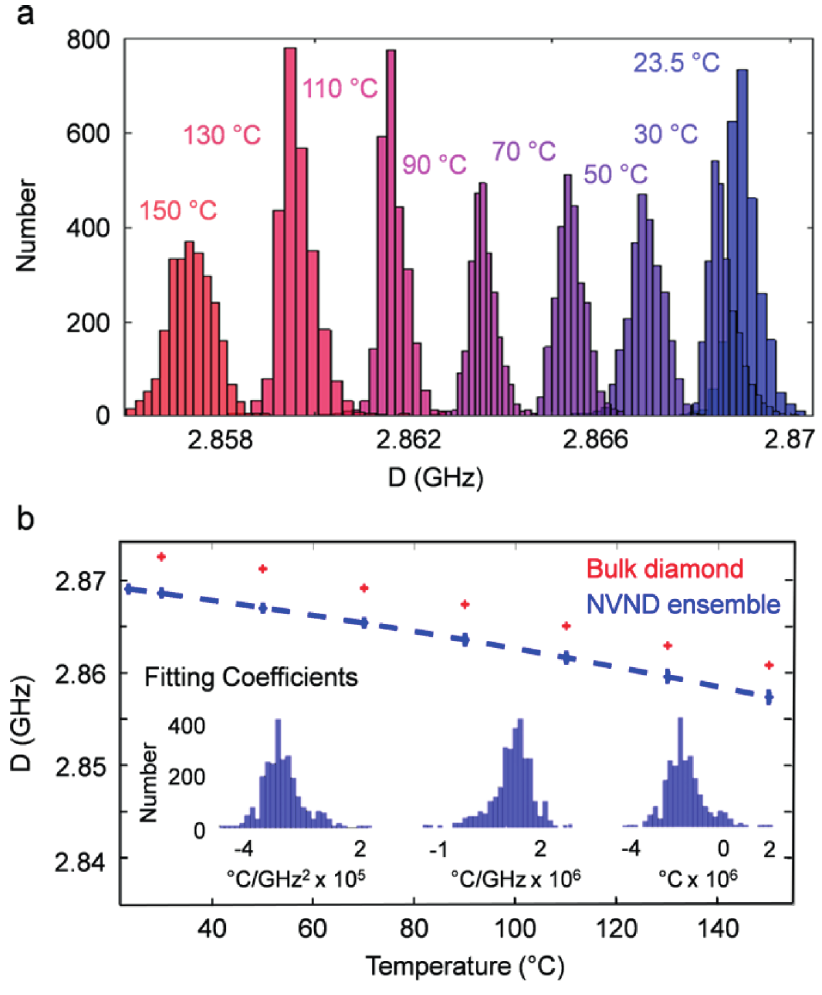}
 \caption{(a) Histograms of $D$ as a function of temperature for NVND ensembles. (b) Mean value of $D$ as a function of temperature for the NVND ensemble and bulk diamonds. (Inset) Histograms of the fitting coefficients (adapted from \cite{foy2019wide}).}
 \label{fig-dDdT}
\end{figure}

\subsubsection*{\textsf{Introduction}}
The precise determination of the temperature dependence of the ODMR shift in spin-based thermometry or the spectral shape in all-optical thermometry is critical for accurate temperature measurements. 
The NDs have nonnegligible particle inhomogeneities, resulting in different temperature responsivities.
In addition, it is challenging to calibrate the absolute value of the temperature dependence. 

\subsubsection*{\textsf{Particle inhomogeneity}}
For spin-based thermometry, the temperature dependence of ZFS ($dD/dT$) was reported to be $- 74$ kHz/K for the NV centers in the bulk diamonds, with a sample variation from $-71$ to $-84$ kHz/K~\cite{PhysRevLett.104.070801,PhysRevX.2.031001,doi:10.1063/1.3652910}. 
Subsequently, $dD/dT$ was found to show greater inhomogeneity in NDs because of crystal strains and surface states~\cite{foy2019wide,Plakhotnik2014all}. 
Foy et al. measured $dD/dT$ for few hundreds of NDs deposited on electronic circuits, and they determined that $dD/dT$ can be described with a second-order polynomial in the range from 23.5--50 $\si{\degreeCelsius}$ (see Fig.~\ref{fig-dDdT}); $dD/dT = -76$ kHz/K for 20--30 $\si{\degreeCelsius}$ and $dD/dT = -115$ kHz/K for 140--150 $\si{\degreeCelsius}$ with a standard deviation of 3.88 K (287--446 kHz using the above two representative $dD/dT$).
To overcome the particle variations, it seems necessary to calibrate $dD/dT$ \textit{in situ} prior to the target measurement, as performed in the case of other fluorescence nanothermometers~\cite{okabe2012intracellular,10.1371/journal.pbio.2003992,rendler2018fluorescent}. 
Further, this strategy for particle calibration is expected to be common to all types of ND family, such as critically magnetic coupled NDs~\cite{PhysRevX.8.011042} and chemically functionalized NDs~\cite{zhang2018hybrid,C9NR03710A}, because these additional coupling components likely increase the inhomogeneity in $dD/dT$. 

In addition to $dD/dT$, the absolute value of ZFS ($D$) may be important for the determination of absolute temperature. 
Although diamond quantum thermometry has not been used for absolute temperature determination, the absolute temperature is of great importance in real applications. 
In principle, the temperature-dependent key properties of the colour centres, ZFS of NV spins in the spin-based thermometry, and the fluorescence spectral shape in the all-optical one, can determine the absolute temperature because they are material properties. 
It is, however, practically challenging in NDs because crystal strains and surface effects create substantial variations in these properties. 
The ZFS of the NV centres can vary by hundreds of kHz, considering the reported spin-Hamiltonian E term of about 100 kHz in highly pure and low-strain diamonds~\cite{PhysRevB.93.024305} (to date, the variation of D has not been reported explicitly).
These excellent NV properties cannot be realized in NDs with the present technology. 
The use of magnetic criticality may be used for absolute temperature determination because it creates a striking abnormality in the ODMR spectral shape caused by the phase transition of the magnetic nanoparticles~\cite{PhysRevX.8.011042}. However, the transition temperature is dependent on the composition of the magnetic nanoparticles (copper and nickel), and achieving composition uniformity in nanoparticles has been a long-standing problem in nanoparticle synthesis technology.
In the all-optical technique, the fluorescence spectral shape is highly dependent on the particles. 
For example, the fluorescence spectra contains information on the different charge states of the defect centres, like NV$^0$ and NV$^-$. 
In addition, the spectral shape can be affected by various optical conditions, such as the transmission of biological samples. 
The determination of absolute temperature by diamond quantum thermometry is yet to be explored.

\subsubsection*{\textsf{Calibration difficulty}}
Besides material inhomogeneity, calibration difficulties can cause variations. 
When measuring $dD/dT$, the ODMR shift is measured by varying the sample temperature, and it is therefore necessary to determine this sample temperature correctly. 
For example, in Ref.~\cite{fujiwara2020realtime}, the ND temperature was assumed to be the surface temperature of the coverslip on which they were spin coated, and this surface temperature was measured using a thin-thermistor tightly attached to the coverslip.
Although this method is a standard technique for measuring surface temperature, the resultant measured surface temperature does not necessarily provide the exact temperature which NDs feel on the coverslip surface. 
The reported value of $dD/dT$ was indeed relatively smaller than the standard value of $dD/dT = -74$ kHz/K because NDs were in contact with air at a relatively lower temperature. 
The $dD/dT$ values were closer to the standard value when the experiments were performed such that the temperatures of the ND substrate and environmental air were set equal~\cite{foy2019wide}. 
In addition to the dependency on the experimental setup, the method of spin-based thermometry affects $dD/dT$. 
In Ref.~\cite{PhysRevResearch.2.043415}, the $dD/dT$ values obtained in the three-and four-point ODMR methods for the same single NDs differ from each other by more than 40\% (three points: $-95$ kHz/K, four points: $-54$ kHz/K). 
The same issue exists in the all-optical thermometry. Choi et al. introduced a multiparametric analysis, using a two-peak Lorentzian fitting function, to the fluorescence spectral shape of the SiV centres~\cite{doi:10.1021/acsphotonics.9b00468}. The challenge here is to identify the temperature range in which the fitting approximation can hold. 
As the temperature varies, the spectral shape of the solid-state emitters deviate, and this may require more Lorentzian peaks in the fitting model. 
It is still unknown how to quantitatively compare the results of diamond quantum thermometry, reported in different papers.

\begin{figure*}[t!]
 \centering
 \includegraphics[width=165mm]{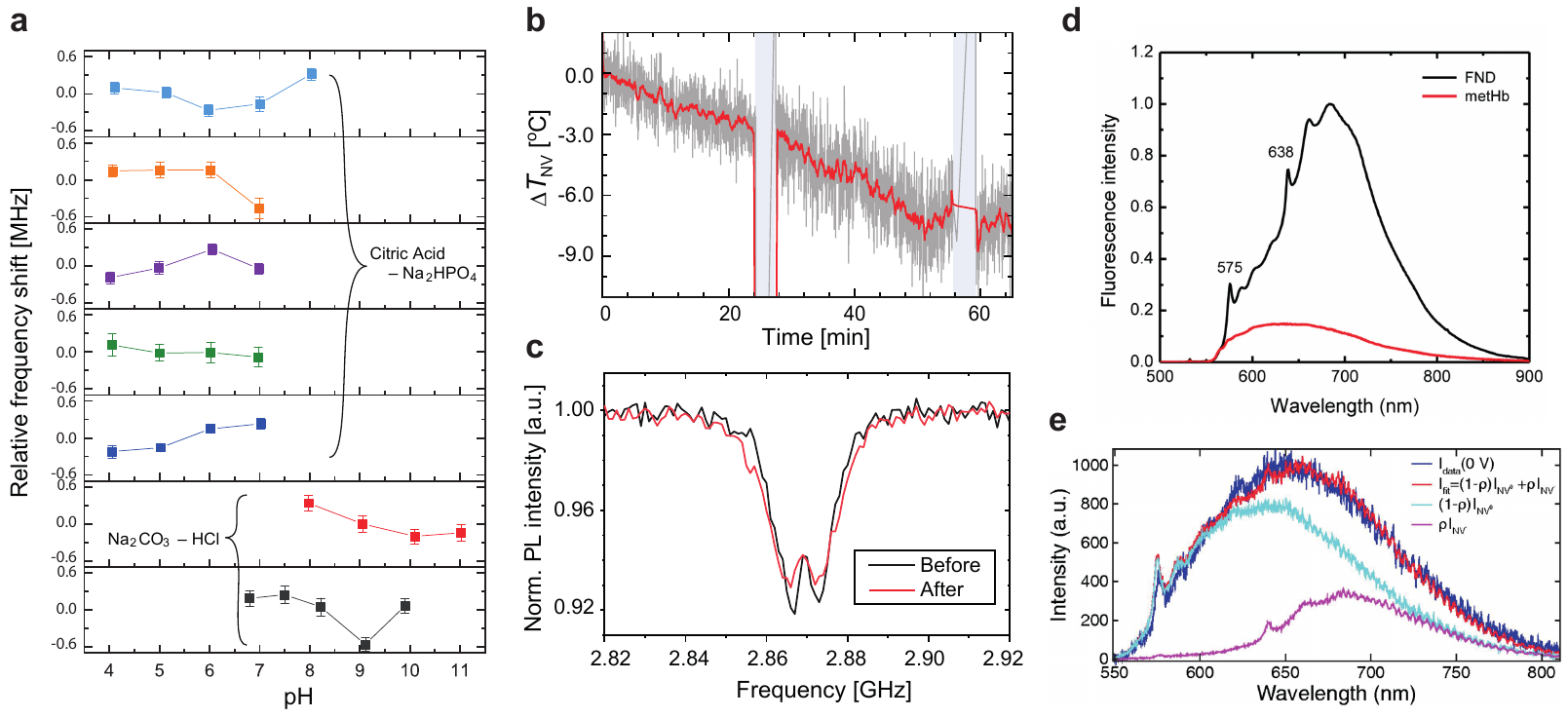}
 \caption{(a) Fluctuation of the ODMR frequency of the NV centres in NDs as a function of pH measured in the CW-ODMR method of spin-based thermometry. The error bars represent fitting errors~\cite{fujiwara2019monitoring}.
  (b) Time profile of $\Delta T_{\rm NV}$ over 70 min measured in the multi-point ODMR method using NV centres and (c) the corresponding CW-ODMR spectra before and after the baseline drift. The blue regions in \textbf{b} indicate periods during which temperature measurements were not performed~\cite{fujiwara2020realtime}.
  (d) Comparison between the fluorescence spectra of fluorescent ND (FND) and methemoglobin (metHb), excited at 532 nm, in water~\cite{Hui2014}.
  (e) Fluorescence spectra of NV$^0$ (cyan) and NV$^-$ (violet) in NDs immersed in a voltage-controllable electrolyte. The real ODMR spectrum from single NDs (blue) consist of both profiles, summed up at a certain percentage (red)~\cite{karaveli2016modulation}.
  }
 \label{fig-pHdep}
\end{figure*}

\subsection{Sensing selectivity and artefacts}
\label{sec3-3}

\subsubsection*{\textsf{Introduction}}
Diamond colour-defect centres exhibit stable spin and optical properties and are considered as robust sensors less likely to generate sensing artefacts in complex environments such as biological samples~\cite{zhang2018hybrid}. However, recent studies have revealed the necessity of experimentally quantifying this sensing robustness~\cite{PhysRevResearch.2.043415,fujiwara2019monitoring,sekiguchi2018fluorescent}. In complex environments, various factors that were not considered to affect the temperature measurements were found to influence the sensing results. 

\subsubsection*{\textsf{Magnetic field}} \ 
Because NV centres are a spin-1 system, the magnetic field is the most influential factor that causes a frequency shift via the Zeeman effect in the diagonal component and spin mixing by the transverse component. 
The diagonal magnetic field lifts the degeneracy of $\ket{\pm 1}$ and can be discriminated from the temperature-dependent ODMR shift if the magnetic field is static beyond the measurement time. 
Therefore, the geomagnetic field does not apparently induce artefacts in spin-based thermometry. Further, it does not affect all-optical thermometry.
The transverse magnetic field can affect both spin and optical properties via spin mixing because it affects the ODMR shift, linewidth, and fluorescence brightness, as observed in colour centres in diamonds and silicon carbide ~\cite{PhysRevA.94.021401,Tetienne2012,PhysRevX.6.031014,doi:10.1063/1.5037158,Simpson2016,Anisimov2016}. SiV, GeV, and SnV are not substantially affected by the magnetic field at room temperature.

\subsubsection*{\textsf{Stress and electric field}} \ 
Stress can affect the spin and optical properties of colour centres in diamonds through the perturbation of the crystal field.
In spin-based thermometry, the pressure change causes an ODMR shift of 15 kHz/MPa~\cite{doherty2014electronic,cai2014hybrid} (0.2 K/MPa), which can be an artefact when applying the thermometry in a dynamic high-pressure environment, such as in high-pressure physics~\cite{Hsieh1349} and aerospace applications~\cite{mi9060276}.
In the case of biological applications, stress is unlikely to affect the temperature measurements because biological pressures are considerably small.

The electric field can cause a frequency shift of ODMR or fluorescence spectral change~\cite{dolde2011electric,iwasaki2017direct}. 
This shift appears in the same direction of the magnetic field Zeeman shift, and the temperature measurement may not be affected, if both ODMR dips of $\ket{0} \rightarrow \ket{\pm 1}$ are probed simultaneously.
In biological applications, there is a proposal to measure local electric fields in a biological context such as the trans-membrane electric field generated by the cell-membrane potential~\cite{schirhagl2014nitrogen}. 
A measurement artefact can arise when biomolecules, possessing large electron affinities, are adsorbed on the ND surface~\cite{doi:10.1002/anie.201905997,doi:10.1246/cl.141036,lin2015protein}. 
Further, charge state conversion caused by the presence of electric fields can become an artefact~\cite{doi:10.1063/1.5139256,karaveli2016modulation}, as we discuss below in the context of pH. 
The effects of such complex scenarios of strong electric fields on diamond quantum thermometry have not been well investigated.

\subsubsection*{\textsf{pH, ions, and water adsorption}}
The effects of pH, ions in the environment, or water adsorption to the surface of diamonds can be clubbed under the family of surface effects on the diamond colour centres, and they can strongly affect both spin-based and all-optical thermometry. 
The surface termination of diamonds affects the spin and optical properties of NV centers~\cite{PhysRevB.83.081304,Grotz2012,kaviani2014proper,Schreyvogel2015,Newell2016surface,karaveli2016modulation,kawai2019Nitrogen,Lofgren2019thebulk}. 
For example, oxygen or nitrogen termination provides positive electron affinity, which makes the NV centres impervious to ionization and associated spin dephasing, whereas hydrogen termination provides negative electron affinity, which is more likely to convert the spin-active negative charge state (NV$^-$) to the inactive neutral state (NV$^0$) (see Fig.~\ref{fig-pHdep}e).
Indeed, the use of ND-NV centres as pH sensors has been demonstrated both in spin-based~\cite{fujisaku2019pH} and all-optical methods~\cite{karaveli2016modulation}. 
The fluctuations in the NV charge states affect both the spin and optical properties of the NV centres, which can become artefacts in the associated thermometry ~\cite{karaveli2016modulation,Yamano2017Charge,Stacey2019evidence,PhysRevLett.122.076101,fujisaku2019pH}.

Artefacts derived from this surface effect may be significant, particularly in biological applications, because of the complexity of the diamond environment in cells and organisms. 
For instance, NDs experience a substantial change in pH in cells after cellular uptake; endosomes exhibit a pH change from 7.0 to 4.0 as they move from the membrane to the nucleus~\cite{Sorkin2002}. 
Thus, the quantification of the level of artefacts is critical for proper interpretation of the diamond quantum thermometry results.
Fujiwara et al. evaluated the immunity of the ZFS ($D$) and $T_2$ coherence time of single NV spins towards pH change over a wide range of pH (4--11), using the spin-based CW-ODMR method~\cite{fujiwara2019monitoring} (see Fig.~\ref{fig-pHdep}a).
Sekiguchi et al. confirmed the immunity of $dD/dT$ to ionic solutions and water~\cite{sekiguchi2018fluorescent}. 
However, these confirmation studies used the CW-ODMR method that provides a precision of approximately 300 kHz (i.e., 4--5$\si{\degreeCelsius}$); 
the immunity still needs to be further confirmed at a higher precision at which the multipoint ODMR method operates (smaller than 10 kHz) and for various surface conditions in diamonds. 
The drift of $D$ was indeed observed at this level using the four-point method (Fig.~\ref{fig-pHdep}b,c). 
While the surface effect on all-optical thermometry has not been clarified very well, the surface-chemistry-dependent background fluorescence in detonation NDs~\cite{Reineck2017effect} can become an artefact in the all-optical method. 

\subsubsection*{\textsf{Microwave and optical heating}}
In spin-based thermometry, microwave irradiation can cause a change in the temperature during the measurement process because of microwave-induced water heating, resulting in an ODMR shift~\cite{An_2020,10.3389/fphy.2020.522536,doi:10.1021/acs.nanolett.8b00895}.
The excitation microwave power depends on the antenna structure and coupling efficiency between the antenna and delivery waveguide. 
In most of the previous spin-based thermometry experiments, the antennas constantly irradiated by microwave radiation and the relative temperature changes were observed in a small region of the order of tens of micrometres; reportedly, the temperature data was not affected considerably. 
However, local heating caused by the conformational changes in biological samples may change the local dielectric permeability, which can affect the local heat-generation rate and temperature. 
Optical absorption is a general challenge for all types of nanothermometry, including both spin-based and all-optical thermometry. 
The optical heating effect is described in detail in Ref.~\cite{bednarkiewicz2020standardizing}.
A shorter excitation wavelength causes a greater heating effect because of the optical absorption characteristics of the biological body.
Such an optical heating effect was studied in detail previously by several authors~\cite{bednarkiewicz2020standardizing,Zhou2020}.

\subsubsection*{\textsf{Biomolecule adsorption}}
The adsorption of biomolecules to the ND surface influences the temperature measurement in biological applications. 
Various biomolecules including proteins adsorbed to NDs in biological environments have a substantial effect on their emission properties and colloidal stability. 
The most evident effect is that their background fluorescence overlaps with the emission of defect centres in diamonds~\cite{doi:10.1063/1.3544312} (see Fig.~\ref{fig-pHdep}d). 
This background fluorescence often exhibits an uncontrollable temporal change in intensity and spectral shape under optical illumination, and they can become artefacts in all-optical thermometry; however, it does not affect spin-based thermometry (background fluorescence diminishes the ODMR contrast but does not cause the ODMR frequency shift). 
The formation of protein corona around the NDs should be prevented using surface functionalization. 
The PG functionalization suppresses protein corona formation in human plasma~\cite{zou2020Polyglycerol,ZOU2020395}, and such surface functionalization may be the key for eliminating artefacts in diamond quantum thermometry.

\subsubsection*{\textsf{Nanodiamond motions and fluorescence intensity fluctuation}}
When NDs are not immobilized in the thermometry experiments, the motions of NDs may cause substantial artefacts in both spin-based and all-optical measurements similar to those caused by other real-time high-speed fluorescence measurements of moving animals~\cite{NguyenE1074}.
In addition to this typical artefact in fluorescence imaging, there is a diamond specific artefact caused by the optical/microwave power dependence of the ODMR spectra in spin-based thermometry. 
Recently, it has been reported that the optical and microwave power dependence of ODMR spectra cause spectral shape variation and frequency shift of ODMR~\cite{PhysRevResearch.2.043415}.
This optical power dependence is probably derived from the interplay between the NV spin coherence time and the optical and microwave decoherence. 
This artefact needs to be treated appropriately when the excitation intensity of NDs can vary, such as during the confocal excitation of mobile NDs under Brownian motion or during the variation of the optical transparency of the ND's environment. 
Such an effort has been reported in the context of spin-based thermometry using NV centres~\cite{Konzelmann_2018}.
Notably, the variation in the optical excitation intensity can also be caused by the optical transparency of the environment around the NDs, such as tissue structural changes, which can result in artefacts both in the spin-based~\cite{PhysRevResearch.2.043415} and all-optical thermometry~\cite{bednarkiewicz2020standardizing,Zhou2020}.

\section{Applications to materials and device physics}
\label{sec4}

\begin{figure*}[t!]
 \centering
 \includegraphics[width=170mm]{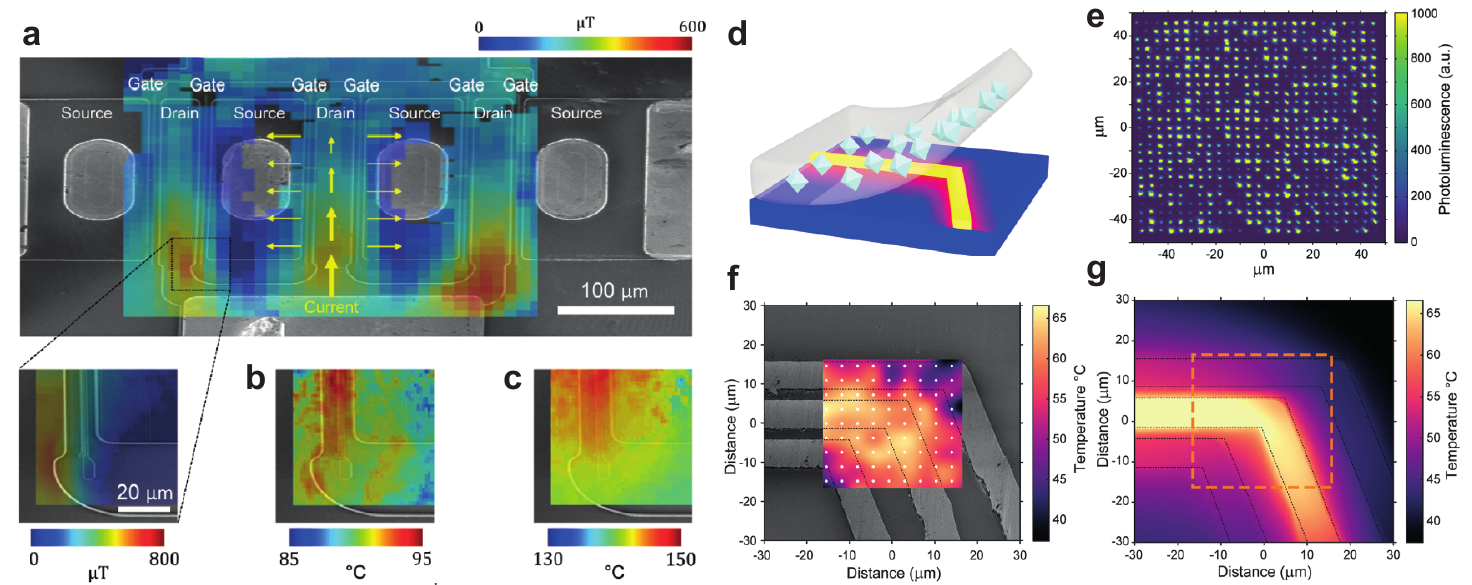}
 \caption{(a) High-resolution magnetic field and (b,c) temperature images of the channel stop (1 W and 1.73 W for drain bias) in a multifinger GaN HEMT (from ~\cite{foy2019wide}). (d) Schematic of the temperature measurements of the heat source coplanar waveguide, patterned on a substrate. (e) 2D fluorescence scan of an ND array in a PDMS matrix (25 $\times$ 25 array of 1000 nm diameter spots with 4 $\si{\um}$ spacing). (f) Temperature map of the waveguide, obtained from interpolated data collected on a 9 by 9 array of NDs (real positions indicated by white dots). (g) Finite-element method simulation for the corresponding region. From~\cite{doi:10.1021/acs.nanolett.8b00895}.}
 \label{fig-edevice-1}
\end{figure*}

\subsection{Introduction}
\label{sec4.1}
Nanoscale temperature measurement is of great importance in modern electronics because of the substantial scaling down of electronic components into the nanoscale. 
At the scientific research level, precise determination of temperature distribution is key to understanding number of thermoelectric phenomena including Joule heating, Peltier effect, and Seebeck effect. 
For industrial applications, Joule heating prevents stable device operation, and thermal management in the system architecture is required. 
However, conventional thermometers are incapable of measuring precise temperature at the nanoscale. 
For example, NIR thermography, which is the most conventional contactless method, relies on the detection of near-infrared thermal radiation with wavelengths of  3--5 $\si{\um}$ or 8--14 $\si{\um}$; it is incapable of spatially resolving objects smaller than these wavelengths. A variety of new thermal measurement techniques that can visualize sizes smaller than several micrometres, such as scanning thermal microscopy and thermo-reflectance microscopy, have been proposed.

Scanning thermal microscopy is a contact method that provides high-resolution down to tens of nanometres. 
This method can be used to visualize the temperature distribution in Ge-Al composite nanowires with a width of only 35 nm under Joule heating and Peltier effect~\cite{gachter2020spatially}. 
Further, this method was used for analysing a complex thermoelectric phenomenon at metal-graphene contacts, and it revealed the roles of Joule heating, Peltier effect, and current crowding~\cite{Grosse2011}. 
Thermo-reflectance microscopy is a contactless method that is useful for characterizing the temperature distribution and heat transport dynamics. 
It provides a spatial resolution of 200--300 nm with a temperature resolution of 10 mK~\cite{Mayer:07}. 
Favaloro et al. used this technique to visualize the Peltier-induced temperature distribution in single crystalline VO$_2$ nanobeam structures with one-dimensional metal-insulator domains~\cite{Favaloro2014direct}. 
This technique can be used to analyse the heat transport dynamics by measuring the thermal conductance directly~\cite{Sood2018,Olson2019spatially}.
Although it has a limited spatial resolution, the lock-in NIR thermography has recently revealed a number of spin-related thermoelectric phenomena, such as the anisotropic magneto-Peltier effect~\cite{Uchida2018,Wang2020} and the spin Peltier effect~\cite{Daimon2016}, by exploiting its high sensitivity of 0.3 mK/$\sqrt{\rm Hz}$~\cite{doi:10.1063/1.1310345}. Motivated by these applications, the last several years have witnessed the emergence of the applications of diamond nanothermometry in electronic devices.

\begin{figure*}[th!]
 \centering
 \includegraphics[width=130mm]{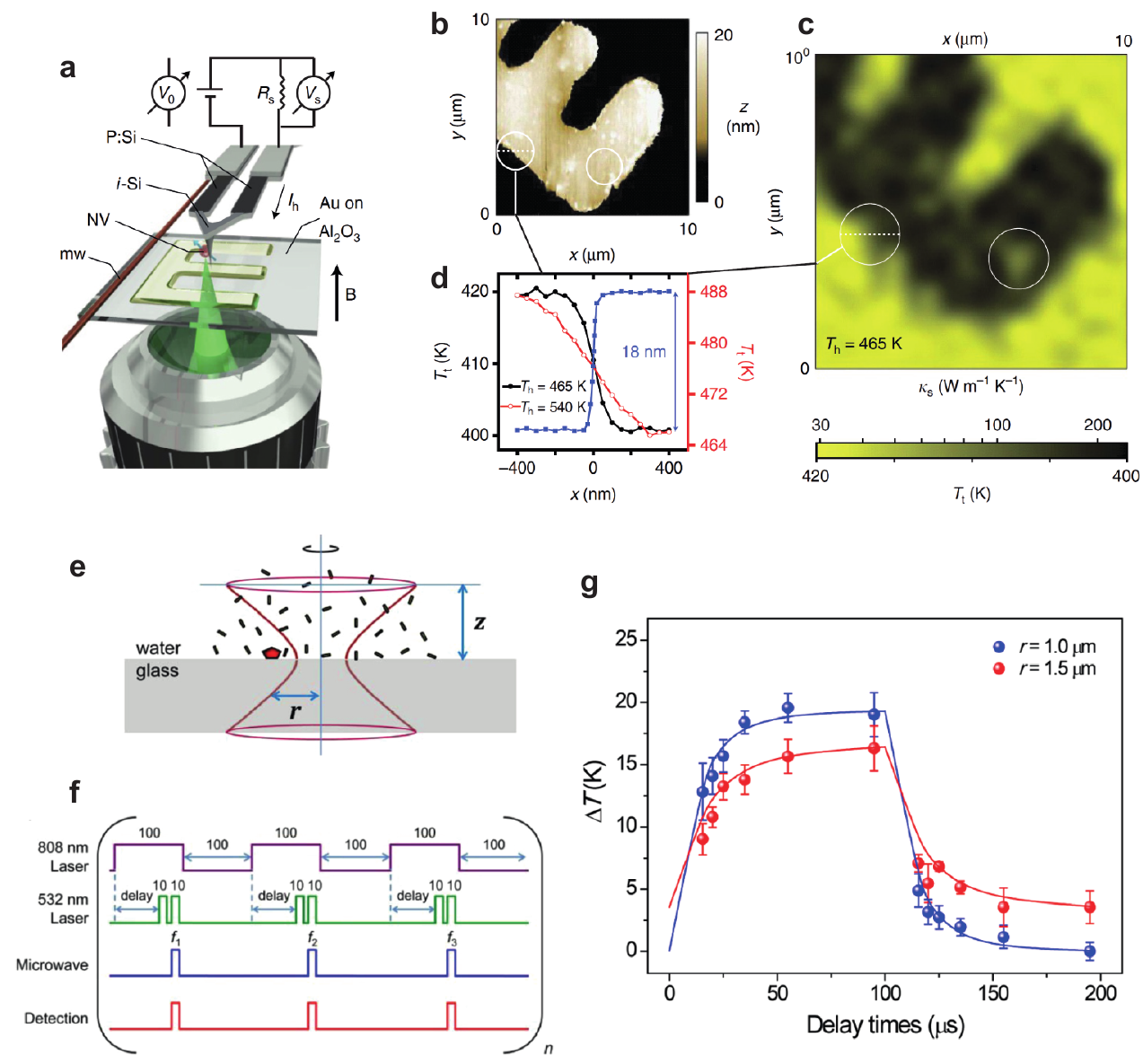}
 \caption{(a) Schematic of a ND-grafted cantilever, scanning over the metallic microstructures, with the spin-based ND thermometry setup. (b) AFM image of the structure and (c) the corresponding NV-assisted thermal conductivity image. 
 The heater temperature $T_{\rm h}$ is 465 K. $T_{\rm t}$ is the tip temperature. The white circles in (b, c) indicate examples of low (high)-conductivity patches, inside (outside) the structure, that do not correlate with the substrate topography. 
 (d) $T_{\rm t}$ near the edge at two different $T_{\rm h}$. The blue curve displays the topography of the film edge (From \cite{Laraoui2015}).
 (e) Schematic of ND thermometry under 808 nm laser heating at the interface of water and glass. Gold nanoparticles were suspended in water; the laser beam focus is denoted by $r$. 
 (f) Timing sequences of the 532 nm and 808 nm laser irradiations, microwave excitation, and fluorescence detection for the time-resolved temperature measurement. All numbers are in units of microseconds. (g) Time evolution of heat dissipation, when irradiated with a pulsed 808 nm laser at $r$ = 1.0 and 1.5 $\si{\um}$. 
 (From \cite{tzeng2015time}).}
 \label{fig-econductance-1}
\end{figure*}

\subsection{Temperature measurement of nanoscale electronic devices}
\label{sec4.2}

Since its invention, diamond quantum thermometry has been used to detect temperature distribution around metallic wires~\cite{neumann2013high}. 
The high controllability of Joule heating and a relatively static experimental platform have enabled various proof-of-principle demonstrations of diamond quantum thermometry.
Neumann et al. reported a sensitivity of 130 mK/$\sqrt{\rm Hz}$ with a spatial resolution of the optical diffraction limit (ca. 200 nm) and an experimental precision of approximately 1--2 K. 
The spatial resolution may be further improved using super-resolution techniques~\cite{Rittweger2009,Barbiero2017}, and the precision may be enhanced with a more sophisticated pulse sequence of the pulsed ODMR.

Foy et al. demonstrated full spatial temperature mapping of joule heating in microfabricated metallic lines with the simultaneous imaging of the magnetic field from the electric current by coating NDs on the circuit structures. 
They employed the CW-ODMR method in a camera-based wide-field technique and analysed the temperature dependent unidirectional ODMR shift and magnetic field driven Zeeman splitting simultaneously.
They further extended this analysis technique to more industrially important applications such as imaging gallium nitride high-electron mobility transistors (GaN HEMTs) that contain a junction between two materials with different band gaps. 
They successfully imaged the magnetic field at the kink of the drain and lower temperature at the end of the gate along the channel direction (Fig.~\ref{fig-edevice-1}a--c), thereby indicating limited leakage current at the end of the channel. 
While the temperature precision of this measurement scheme was a few Kelvin owing to the use of the CW-ODMR method, it was comparable to micro-Raman thermography~\cite{974795,1705092} and can be refined using multi-point ODMR or pulsed ODMR techniques.
Recently, determination of the Joule heating around metallic wires via all-optical diamond quantum thermometry has also been reported~\cite{Hui2019}.

Another approach for obtaining full spatial temperature mapping is to fabricate microarrays of NDs into polymer film sheets. 
Andrich et al. fabricated an array of NDs on the film sheet of polydimethylsiloxane (PDMS)~\cite{doi:10.1021/acs.nanolett.8b00895} (Fig.~\ref{fig-edevice-1}d--e). 
This ND-array PDMS sheet was transferred to coplanar waveguides, where the heating microwave radiation was applied as input. 
They visualized the temperature distribution around the waveguide. 
While this technique needs to consider the heat capacity and conductivity of polymer film sheets to measure the temperature correctly, it has the potential to use the nanoparticulated bulk-grade NV centres for thermometry~\cite{doi:10.1021/nl501208s}, which significantly improves the temperature resolution to several milli-Kelvin. 

\subsection{Nanoscale thermal conductance measurements}
\label{sec4.3}
Laraoui et al. visualized the thermal conductance of metallic microstructures using NDs attached to the AFM tip~ \cite{Laraoui2015} (Fig.~\ref{fig-econductance-1}a--d). 
In this experiment, the AFM cantilever had electrical lines to generate heat, and the tip temperature in the range of 300--560 K was measured with NDs in the CW-ODMR method. 
As the ND-grafted AFM tip was scanned over metallic microstructures, the tip temperature decreased depending on the thermal conductivity of the metallic microstructures, thereby visualizing thermal conductivity. The sensitivity for the thermal conductance was 5\%/$\sqrt{\rm Hz}$, and it anticipated 0.2 \% /$\sqrt{\rm Hz}$.
They determined 184 $\si{\us}$ as the characteristic time response constant for their particular application, while the ultimate time response was limited by the NV spin readout time of $\sim 1 \ \si{\us}$.

The potential for the fast temperature measurements down to the microsecond time scale was studied by Tzeng et al. ~\cite{tzeng2015time} (Fig.~\ref{fig-econductance-1}e--g), although their demonstration was not aimed at electronic device applications. 
They proposed a three-point ODMR method in spin-based thermometry using NV centers and combined it with laser heating to investigate thermal dynamics of gold nanoparticles under a confocal fluorescence microscope. 
By laser heating a radius of 1 $\si{\us}$, a thermal response with a characteristic time of $12 \ \si{\us}$ was obtained.
The measurement of transient temperature dynamics, under a microscope, by the time-domain thermo-reflectance (TDTR) method has been studied extensively as a conventional technology~\cite{Olson2019spatially,doi:10.1063/1.5046944}.
This method is based on the ultrafast pump-probe spectroscopy to monitor the acoustic waves and analyses the heat transport dynamics. While TDTR method technically limits the delay time, the diamond quantum thermometry may be applied to the long-time heat transport dynamics.

\section{Applications to biological science}
\label{sec5}

\subsection{Introduction}
\label{sec5.1}
Temperature regulates the rate of chemical reactions in molecular systems and is a crucial parameter for cellular activities~\cite{doi:10.1146/annurev.ph.57.030195.000355}. In thermal biology, the physiological properties are revealed by measuring the temperature in biological systems~\cite{TB0,TB1,TB2,TB3,TB4,TB5}. To understand the hierarchical gap between molecular states and biological functionalities, dynamical behaviours and states of the biological systems need to be understood. This requires measuring the nanoscale temperatures with high precision.

Diamond quantum thermometry is one of the candidates for measuring temperatures by local florescence thermometers~\cite{sotoma2017review}. This method has been applied to various cultured cells, including cell-line cells, neurons, and stem cells to validate the real temperature changes inside the cells. One of the motivations for determining the internal temperature of cells is to interpret the results of fluorescent nanothermometry of living cells. Numerous studies have shown that the intracellular temperature is higher than the environmental temperature (or temperature heterogeneity), whereas the thermometry results, yielding a macroscopic physical estimation based on the heat dissipation and calorimetry results, contradict this observation.

The other motivation is to determine the real temperature of cells to obtain a correlation between the cellular temperature and cellular functions. This is important for the recent technical developments in tissue engineering and regenerative medicine because temperatures at which cells grow in these applications are temporally variable and spatially heterogeneous.

\subsection{Local temperature probing under photothermal laser heating}
\label{sec5.2}

\begin{figure*}[th!]
 \centering
 \includegraphics[scale=1.0]{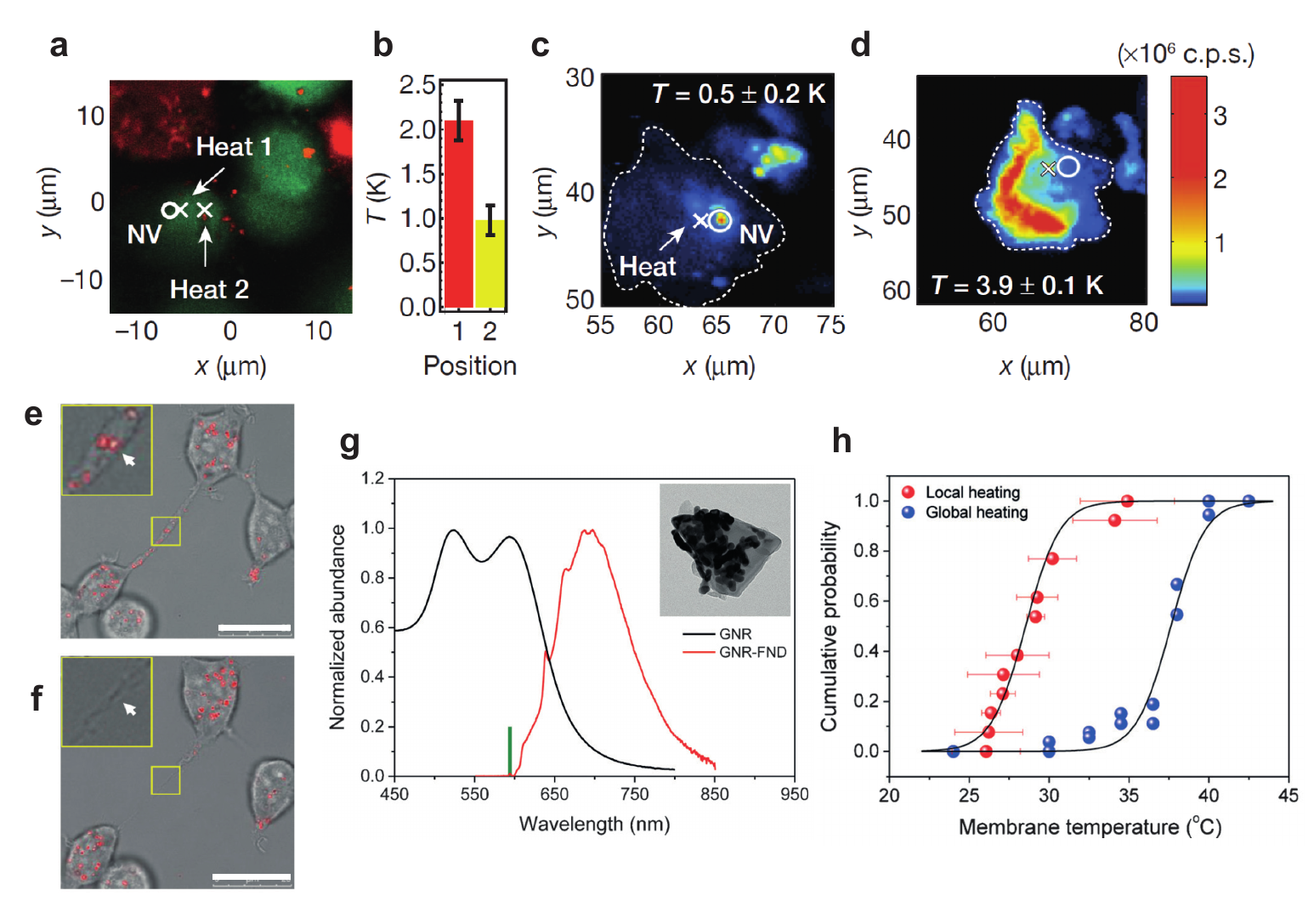}
 \caption{(a) Fluorescence scan of stained cells (green; live cells. red; dead cells). (b) Temperature of a single ND (circle in \textbf{a}) with local heat applied at two different positions (crosses in \textbf{a}, Heat 1 and Heat 2). 
 (c) Fluorescence scans of an individual cell under low-intensity heating laser corresponding to the increase of 0.5 K and (d) 3.9 K. 
 Cell death is indicated by the penetration of staining dye into the cells. From \cite{kucsko2013nanometre}. 
 (e) Merged bright-field/fluorescence (right) images of the HEK293T cells labelled with GNR-FNDs (red). The cells are connected via membrane-tunnelling nanotubes (TNTs). 
 (f) Corresponding images after exposure to heating laser (TNT ruptured). 
 The insets show the regions with laser-heated particles in the yellow boxes (while arrows: target particles). Each scale bar indicates 25 $\si{\um}$.
 (g) Absorption spectrum of GNRs (black) and the emission spectrum of GNR-FNDs (red). The 594 nm excitation laser is indicated as the thick green bar. Inset depicts TEM image of a GNR-FND. Typical size of GNR is $10 \times 16$ nm, while that of FND is 100 nm.  
 (h) Empirical cumulative distribution plot of the membrane temperatures at which TNTs are ruptured by the local and global heating (from \cite{tsai2017measuring}).
}
 \label{fig-vitro-heating}
\end{figure*}

\subsubsection*{\textsf{Local temperature probing under photothermal treatment in cultured cells}}

The first demonstration of diamond quantum thermometry for cells was reported by Kucsko et al. in combination with NIR-laser local heating~\cite{kucsko2013nanometre} (see Fig.~\ref{fig-vitro-heating}a--d). 
They achieved a precision of 8 mK using NV spin-based thermometry in pure bulk diamonds. 
After demonstrating the potential temperature precision of spin-based thermometry, they deployed this measurement principle to NDs incorporated into cells and measured the temperature of specific points in human embryonic fibroblast cells with a temperature precision of 0.1 K while creating the temperature gradient inside the cells using laser heating. 
They used the multipoint ODMR method for spin-based thermometry. Further, they continuously monitored the intracellular temperature until cell death was caused by heating. 
This work was significant because it presented the possibility of a drastic improvement in nanothermometry in biological environments using quantum technology; however, the ultimate temperature sensitivity can be achieved only with NV centres in pure bulk diamonds. 

The combination of diamond nanothermometry with local heating was subsequently applied to characterize the thermostability of cell membranes. %
Tsai et al. investigated the critical temperatures for the rupture of membrane nanotubes and photoporation of cell membranes in human embryonic kidney cells using all-optical diamond nanothermometry in combination with laser heating~\cite{tsai2017measuring} (see Fig.~\ref{fig-vitro-heating} e--g). 
They used nanohybrids of NDs and gold nanorods (GNRs) and measured the temperatures of the exact points of excitation. 
In this experiment, the temperature was monitored by the wavelength shift of the ZPLs in the NV fluorescence spectra and all-optical thermometry, and the hybrids were simultaneously heated up by a longitudinal surface plasmon with the same excitation laser. 
The results indicate that these membrane structures are ruptured at considerably lower temperatures than those required for global heating. 
Notably, local heating with photothermal polymer-coated NDs have been recently reported as an alternative to ND hybrids to improve the material controllability and durability~\cite{Sotomaeabd7888,yingkewu2020chemRxiv}.

\subsubsection*{\textsf{In vivo local temperature probing and laser heating for embryogenesis in \textit{C. elegans}}}
Recently, diamond quantum thermometry was used for \textit{in vivo} applications to probe the temperatures inside multicellular model organisms. 
Choi et al. demonstrated the manipulation and probing of local temperature inside embryos of \textit{C. elegans} using NV spin-based thermometry and laser heating (see Fig.~\ref{fig-vitro-heating}e--g). 
They decorated NDs with $\alpha$-lactalbumin/PEG functionalization and introduced them into gonads. 
As the worms matured, ND-injected early embryos were extracted from the worms via dissection and their cell division process, under laser heating at nucleus, was observed. 
They measured the local temperature inside the cell by the multipoint method of spin-based diamond quantum thermometry with an ultrafast particle tracking algorithm that can lock onto the ND particles under Brownian motion~\cite{Fields:12}. 
The results revealed that the cell division time was influenced by the average temperature of the cell and not the nucleus temperature---a long-standing question in developmental biology. 

\begin{figure*}[th!]
 \centering
 \includegraphics[width=170mm]{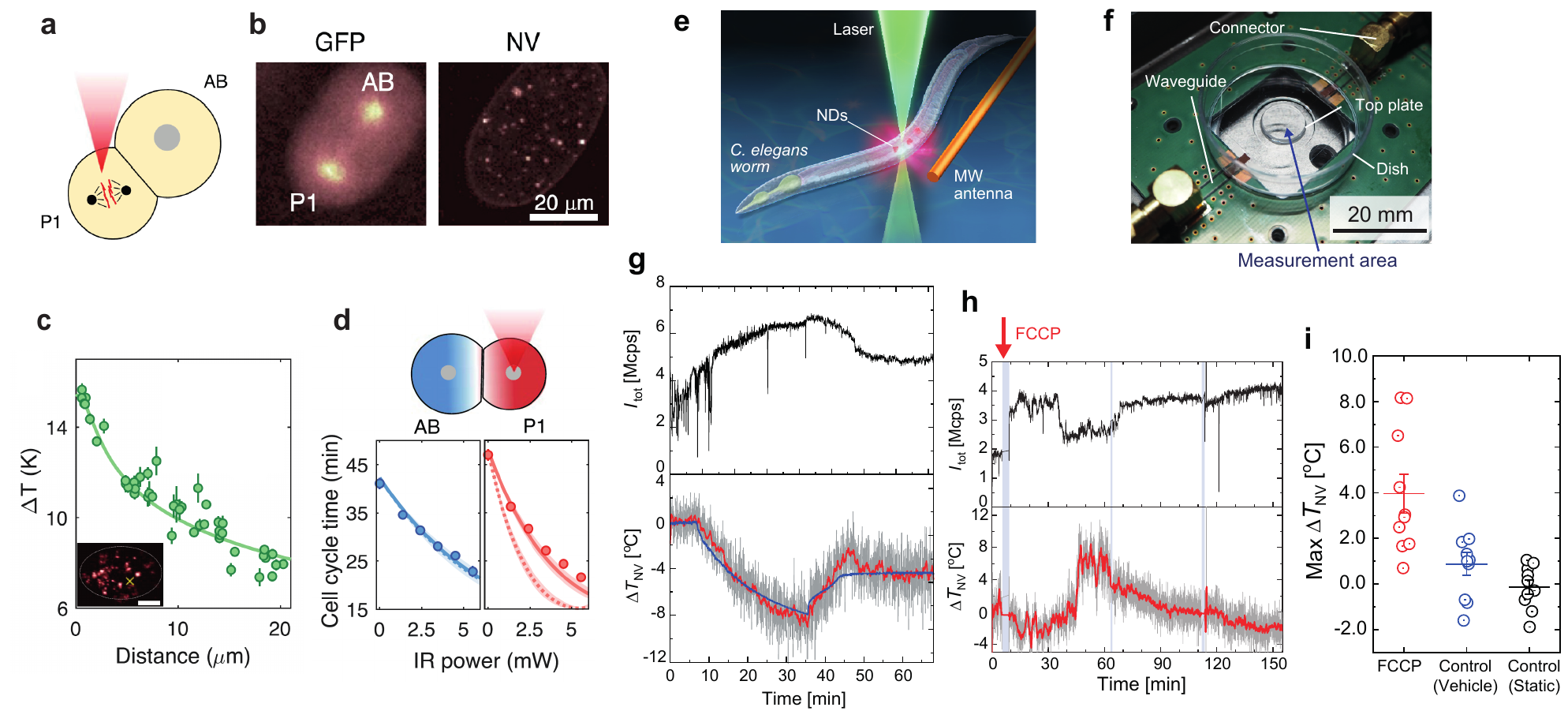}
 \caption{
 (a) Schematic of selective heating of P1 in \textit{C. elegans} two-cell embryos. 
 (b) Confocal images of the GFP-labelled embryo with GFP imaging (left) and NV-ND fluorescence imaging (right). 
 (c) \textit{In vivo} temperature distribution measured by a collection of NV NDs inside an embryo. The solid line denotes the simulated temperature profile. The inset shows a fluorescence image with a yellow cross pinpointing the laser heating position (Scale bar: 20 $\si{\um}$).
(d) Cell-cycle (cell-division) times of the AB and P1 cells, which are the two-cell embryo, as a function of IR power at P1 nucleus. The solid and dashed lines are theoretical predictions based on average and nuclei temperatures of individual cells, respectively. The experimental data is well fitted to the solid lines.
 From~\cite{choi2019}.
 (e) Schematic of NDs probing inside the worms. (f) Photograph of the measurement device. 
 (g) Time profiles of ND fluorescence intensity ($I_{\rm tot}$) (top) and the relative temperature change indicated by the ND thermometers ($\Delta T_{\rm NV}$) (bottom). Gray: $\Delta T_{\rm NV}$ sampled every 1 s. Red: moving average. Blue: environmental temperature change. (h) Time profiles of $I_{\rm tot}$ (top) and $\Delta T_{\rm NV}$ (bottom) during the FCCP stimulation. (i) Statistical plots of the maximum $\Delta T_{\rm NV}$ for FCCP stimulation (red), vehicle-control (blue), and static control (black, no solution added) (From \cite{fujiwara2020realtime}).
 }
 \label{fig-vivo-heating}
\end{figure*}

\subsubsection*{\textsf{Precise photothermal laser heating for thermogenetic control}}
Ermakova et al.~\cite{Ermakova2017} reported another example by developing a fibre-optic probe that integrated NV-incorporated diamond microcrystals at the fibre tip with a microwave circuit. This circuit can be used to target the cells for single-cell temperature measurements.
They aimed to manipulate neural activity via snake-derived thermosensitive transient receptor potential (TRP) cation channels, genetically expressed in cultured hippocampal/neocortical neurons. 
They demonstrated thermogenetic excitation at a single cellular level and performed a \textit{in vivo} demonstration of the thermogenetic control in zebrafish larvae. 
Thus, diamond quantum thermometry contributes to the development of the site-specific control in thermogenetics.

\subsection{Temperature probing for regenerative medicine and pharmacological treatment}
\label{sec5.3}
\subsubsection*{\textsf{Efforts to extend diamond quantum thermometry to tissue engineering and regenerative medicine}}
There is a substantial need to evaluate the real temperature of cells in tissue engineering and regenerative medicine. 
Previous works have focused on developing suitable \textit{in vitro} systems.
Simpson et al. evaluated the cytotoxicity of NDs and demonstrated ND thermometry in primary cortical neurons~\cite{simpson2017non}. 
They applied multielectrode array recordings to cultured neurons and evaluated whether the neuronal responses were affected by the uptake of NDs. 
After determining neuronal nontoxicity, they applied spin-based thermometry using the CW-ODMR method and confirmed that the intracellular temperature varied by $-1.36 \si{\degreeCelsius}$ under a global temperature change of $-2.1 \si{\degreeCelsius}$.
There is another report on the formation of neural networks from a culture of live differentiated neural stem cells with the help of tissue engineering technology, namely the synthetic scaffolds of electrospun polymer nanofibres incorporating NDs~\cite{price2019quantum}. 
Furthermore, these experiments can be extended to tissue slices on multielectrode arrays in combination with thermal stimuli~\cite{10.3389/fnins.2016.00135,Spira2013}.

For applications in regenerative medicine, such as regenerative tissues~\cite{HALPERIN20124975,ARISAKA201697} and organoids~\cite{Gjorevski2016,10.1371/journal.pone.0199412}, Yukawa et al. correlated the intracellular temperature with the production of growth factors for regeneration by means of ND thermometers in adipose-tissue-derived stem cells (ASCs), taken from mice~\cite{yukawa2020quantum}. 
They evaluated the influence of the global culture temperature on the production of growth factors from ASCs. 
They next developed culture dishes with microwave antennas and calibrated the global culture temperature to the intracellular temperature, measured using diamond quantum thermometry. 
Further, they evaluated the cytotoxicity of NDs from multiple aspects of regenerative medicine, such as proliferation and differentiation abilities. 

\subsubsection*{\textsf{Temperature probing of \textit{C. elegans} adult worms under pharmacological treatment}}

\textit{In vivo} applications will require substantial advancement of diamond quantum thermometry techniques. 
Fujiwara et al. developed a real-time temperature measurement system and demonstrated temperature probing inside young adult worms of \textit{C. elegans}~\cite{fujiwara2020realtime}. 
They developed disposable antenna-integrated dishes and integrated particle tracking and error-correction filters that compensated the artefacts, created from the optical power dependency of NV spins, into the algorithm of the four-point method of spin-based thermometry. 
This system enabled stable temperature monitoring inside adult worms under external temperature changes. In addition, they detected a temperature increase in the worm under the chemical stimuli of the mitochondrial uncoupler.
It turns out that this study presents some challenges to diamond quantum thermometry, including the difficulty in determining the precise temperature derived from the ND particle inhomogeneity of $dD/dT$. 
They discussed that, without calibrating $dD/dT$ for each ND before temperature measurement, the temperature indicated by the ND thermometers shows a substantial range of uncertainties.
In addition, the results need to be evaluated in reference to the decade-long debate on the thermogenic temperature rise in living cells or microorganisms~\cite{suzuki2020challenge,Ghonge_2018,baffou2014critique,Suzuki2015}.

\subsection{Combining ND thermometers with other optical nanothermometers and assays}
\label{sec5-3}

Precision and reliability are crucial to nanothermometers employed in biological environments. 
The ND thermometers proved to be robust in various surrounding environments (see Sec.~\ref{sec3-3}), thereby increasing the reliability of diamond quantum thermometry.
Alkahtani et al. reported the simultaneous use of NDs and lanthanide ion-doped upconversion nanoparticles (UCNPs) for double-mode temperature measurements in fertilized bovine embryos~\cite{Alkahtani:17}. 
In UCNP-based thermometry, the temperature can be determined by the ratio of fluorescence intensities at two fluorescence peaks, located at 525 nm and 550 nm, which exhibit the temperature dependence following the Arrhenius law. 
They injected both NDs and UCNPs into the embryos, prior to cell division, and performed ODMR measurements, for the NDs, and fluorescence spectral measurements, for the UCNPs. 
By heating the embryos using a 980 nm laser, they monitored the temperature of the embryos using both thermometers. 
A sensitivity of 0.27 K/$\sqrt{\rm Hz}$ was demonstrated by the ND thermometry and 0.32 K/$\sqrt{\rm Hz}$ by the UCNP thermometry ~\cite{Alkahtani:17}.

For future biological applications, in addition to increasing the reliability of the diamond thermometers, combining them with other biological assays that provide different perspectives to the observed target will be required. 
An example is presented in a previously reported work, wherein the accuracy of fluorescent nanothermometers, used to measure the temperature of mitochondria, increased up to 50 $\si{\degreeCelsius}$~\cite{10.1371/journal.pbio.2003992}.
Although the interpretation of the result indicating 50 $\si{\degreeCelsius}$ accuracy is debatable, this study performed well-designed experiments, including various control experiments for the mitochondrial respiratory chains and developing temperature calibration strategies, thus providing a good example of how diamond thermometers need to be used in future biological experiments.

\section{Conclusion and outlook}
\label{sec6}
Diamond quantum thermometry is recognized as a quantum sensing method to utilize the optical and electronic spin properties of the colour centres in diamonds with high spatial resolution and sensitivity. 
In this review, we recapitulated the fundamental issues related to diamond quantum thermometry, including the measurement principle and material properties, and discussed several technical issues on the robustness of the method and possible artefacts. Furthermore, several case studies in materials and biological sciences are introduced with demand-based technological requirements. From such systematic approaches, diverse practical applications of diamond quantum thermometry in nanoscience are expected to appear in the near future.

Several challenges remain to be addressed for the future development of this technique, although the feasibility of diamond quantum thermometry has been demonstrated for various nanoscale thermometry applications. 
First, for each application, the pros and cons of diamond quantum thermometry method should be compared with the other thermometry techniques. Proper thermometry methods and measurement techniques should be selected to reveal thermal properties and functions in any physical, chemical, and biological systems, especially in the nanoscale. Based on these requirements, the all-optical diamond quantum thermometry has emerged as a suitable thermometry method because of its simplicity and compatibility with other optical thermometry techniques. Indeed, several authors have initiated various studies to compare the different luminescent thermometry techniques for developing a technical standardization in quantum thermometry. 
Second, further technological refinement of diamond quantum thermometry is necessary to detect various interesting thermal phenomena in electronic devices and biological species. In addition to high temperature sensitivity (precision) and high accuracy, artefact-free sensing capability and effective sensor positioning methods are required as well. 
Third, the effective integration of diamond quantum thermometry with a variety of fluorescence imaging and thermal manipulation techniques, such as high-speed imaging, large-area imaging, and nanoscale heating, is required for the diverse applications of nanothermometers. 
In summary, diamond quantum thermometry has the potential to shine as an effective quantum sensing method by taking root in scientific and technological applications. Consequently, it is an emerging field, which can become the calm technology applicable in various scales.

\section*{Acknowledgments}
We thank T. An, H. Ishiwata, and Y. Zou for their helpful discussions.
We acknowledge the funding from the Osaka City University Strategic Research Grant 2017--2020, JSPS-KAKENHI (19K21935, 20H00335), MEXT-LEADER program, JST PRESTO (No. JPMJPR20M4), Murata Science Foundation, Sumitomo Foundation, and Watanabe Foundation.

\section*{References}
\bibliographystyle{iopart-num}
\bibliography{mainbib}
\end{document}